\documentclass[12pt]{article}

\usepackage[totalheight = 23cm, totalwidth = 17cm]{geometry}
\usepackage{amssymb,amsmath,amsfonts,amsbsy,graphicx,bm}

\usepackage{color,cancel,ulem}

\newcommand{\dis}[1]{\begin{equation}\begin{split}#1\end{split}\end{equation}}

\begin{document}

\begin{titlepage}

\begin{center}

{\LARGE \bf 
Information paradox and island in quasi-de Sitter space
}

\vskip 1.0cm

{\large
Min-Seok Seo$^{a}$ 
}

\vskip 0.5cm

{\it
$^{a}$Department of Physics Education, Korea National University of Education,
\\ 
Cheongju 28173, Republic of Korea
}

\vskip 1.2cm

\end{center}

\begin{abstract}

 Whereas a static observer in de Sitter (dS) space detects   thermal radiation emitted by the horizon, the dS isometries impose that the radiation is  in   equilibrium with the background.
This implies that for the static observer to find the information paradox, the background must be deformed to quasi-dS space  in which the dS isometries are spontaneously broken.
 We study the condition that the information paradox arises in  quasi-dS space with the monotonically increasing horizon size  which is used to describe the inflationary cosmology.
 For this purpose,   the dimensional reduction of three-dimensional dS space with thermal radiation modelled by the JT gravity coupled to CFT is considered.
 We argue that when the central charge monotonically increases in time, the information paradox arises but the conditions for the existence of the island become more restrictive.
 As the central charge can be interpreted as the number of degrees of freedom, the absence of the island in quasi-dS space  supports the entropy argument for the dS swampland conjecture.

\end{abstract}

\end{titlepage}

\newpage

\section{Introduction}

 The black hole information paradox \cite{Hawking:1976ra} has been one of the most puzzling issues in quantum gravity over  several decades.
 It arises as the semiclassical  entropy of the Hawking radiation monotonically increases in time, eventually exceeding the black hole entropy given by $S_{\rm BH}={\rm (Horizon~Area)}/(4G)$.
 This is incompatible with  so-called central dogma, which claims that  black hole as seen from the outside is described by the unitarily evolving quantum system with ${\rm exp}[S_{\rm BH}]$ degrees of freedom.
 Recently, a resolution to the paradox at the semiclassical level has been suggested \cite{Penington:2019npb, Almheiri:2019psf, Almheiri:2019hni, Penington:2019kki, Almheiri:2019qdq} (see also \cite{Almheiri:2020cfm, Raju:2020smc} for  reviews).
 It is based on the finding that  the dominant saddle of the Euclidean path integral on the replica manifolds, which is used to calculate the radiation entropy,  includes the black hole interior called an ``island" at late time.
 Then the full entropy of radiation after some time called Page time  is decreased by the entanglement between radiation and island, realizing the Page curve \cite{Page:1993wv, Page:2013dx}, the time evolution of the radiation entropy consistent with the central dogma.

 This remarkable progress in black hole physics motivates the application of the `island rule' to other quantum gravity systems.
 Among possible backgrounds composing the quantum gravity system, de Sitter (dS) space has drawn considerable interest \cite{Chen:2020tes, Hartman:2020khs, Balasubramanian:2020xqf, Sybesma:2020fxg, Geng:2021wcq, Aalsma:2021bit, Kames-King:2021etp, Shaghoulian:2021cef, Teresi:2021qff, Bousso:2022gth, Espindola:2022fqb} since an observer in dS space can access the finite portion of spacetime surrounded by  the event horizon only. 
 The observer cannot receive signals from the region beyond the horizon  due to the accelerating expansion, which leads to the production of thermal radiation near the horizon \cite{Gibbons:1977mu}.
 Since it is quite similar to  the Hawking radiation process in black hole,  one may na\"ively expect that the information paradox  arises in dS space as well.
 However, in spite of several similarities, the thermodynamics of dS space is quite different from that of black hole (see, e.g., \cite{Jacobson:2003wv} for a review).
 The difference mainly comes from the fact that the dS horizon is not a localized object in space, but an observer dependent one.
 The  radiation emitted by the dS horizon will pass by the observer and recede beyond the horizon.
 The dS isometries guarantee that the radiation is emitted and absorbed by the horizon    in all directions at the same rate, resulting in the thermal equilibrium of the radiation with the surroundings.
 Thus unlike the evaporating black hole,  dS space maintains the constant  horizon size and we can find a time direction along which the radiation entropy is static such that the information paradox does not arise in the static patch, the region inside the horizon.
 \footnote{Whereas   black hole formed by the collapse of star evaporates, the eternal black hole maintains the constant size through the balance between the energy flux emitted by the past horizon and that absorbed by the future horizon. 
 The quantum state for the collapsing black hole is called the Unruh state \cite{Unruh:1976db}, whereas that for the eternal black hole is called the Hartle-Hawking state \cite{Hartle:1976tp}. 
 The dS analogy of the Hartle-Hawking state is the Bunch-Davies vacuum in which dS isometries are preserved  \cite{Chernikov:1968zm, Bunch:1978yq}. }

 The situation can be changed in quasi-dS space, in which  some of dS isometries are slightly broken by the background.
 In this case, the energy flux absorbed by the horizon is not balanced with that emitted by the horizon any longer.
 Then the radiation entropy evolves in time, and furthermore,  the background geometry may be deformed by the backreaction of the nonzero net  energy flux \cite{Aalsma:2019rpt, Gong:2020mbn} (see also \cite{Aalsma:2021bit}).
 When the horizon radius of quasi-dS space $H^{-1}$ varies in time,  we can compare the time evolution of the radiation entropy with that of the geometric entropy given by $S_{\rm dS}=({\rm Horizon~Area})/(4G)$.
 An observer in the static patch can find the information paradox if the radiation entropy exceeds the geometric entropy (see also \cite{Kames-King:2021etp} for a relevant discussion).

  Meanwhile, the universe is  believed to have experienced   inflation  at the early stage, which is well described by the quasi-dS background.
  Since the current universe has only a small amount of dark energy giving $H \sim 10^{-60}G^{-1/2}$, we expect that $H$ has decreased, or equivalently, the geometric entropy has increased in time during the inflation.
  Therefore, for the inflationary quasi-dS space to have the information paradox, the radiation entropy needs to increase much faster than the geometric entropy.  
  In this article, we investigate when this condition is satisfied and whether we can find the island in the inflationary quasi-dS space.  
  For this purpose, we consider the two-dimensional Jackiw-Teitelboim (JT) gravity  \cite{Teitelboim:1983ux, Jackiw:1984je} coupled to   conformal field theory (CFT).
  As reviewed in Sec. \ref{Sec:JTgravity}, the JT gravity can be obtained by  the dimensional reduction of   three-dimensional dS space, so we in fact address the information paradox in  the three-dimensional quasi-dS background. 
  Moreover, our discussion is made in terms of the static coordinates, which are appropriate to describe the dynamics of the horizon as seen by an observer in the static patch.
  In Sec. \ref{Sec:dSparadox}, we argue that the  inflationary quasi-dS space can have the information paradox when the central charge increases in time.
  Intriguingly, since the central charge can be interpreted as the number of degrees of freedom,  the appearance of the information paradox in this case can be connected to the entropy argument in \cite{Ooguri:2018wrx} supporting the dS swampland conjecture \cite{Obied:2018sgi, Andriot:2018wzk, Garg:2018reu, Ooguri:2018wrx}.
  That is, as the modulus responsible for the vacuum energy traverses along the trans-Planckian geodesic distance, towers of states descend from UV \cite{Ooguri:2006in}, resulting in the rapid increase in the number of low energy degrees of freedom.
  We note here that the distance conjecture is not so well established in two dimensions \cite{Etheredge:2022opl}.
 Since  two dimensional gravity we consider is obtained by  a  dimensional reduction of three dimensional gravity,   we assume    the increase in the number of degrees of freedom to be an effect of three dimensional gravity.
  In two dimensional CFT perspective, this may not contradict to the Zamolodchikov's c-theorem which states $c_{\rm UV}>c_{\rm IR}$  \cite{Zamolodchikov:1986gt} as the increase in the number of degrees of freedom does not mean that IR degrees of freedom are newly generated but UV degrees of freedom descend.
 Since the distance conjecture assumes a tower of states in UV, we expect many more degrees of freedom still present in UV.

  If quasi-dS space does not have the island, the radiation entropy produced in this way keeps monotonically increasing in time, contradict to the central dogma \cite{Teresi:2021qff}. 
 Then we expect the quasi-dS background cannot persist for arbitrarily long times \cite{Ooguri:2018wrx} (see also \cite{Seo:2019mfk, Seo:2019wsh, Sun:2019obt} for more discussions).
  In Sec. \ref{Sec:island}, we check if the island can exist in the inflationary quasi-dS space when the central charge increases in time.
  For this purpose, we use three conditions in \cite{Hartman:2020khs} which restrict the spacetime region allowing  the island to exist.
  From this we conclude that the island does not exist in the inflationary quasi-dS space thus for the consistency with the central dogma, the background must be strongly deformed, as the entropy argument for the dS swampland conjecture claims.

\section{JT gravity on dS$_2$  coupled to CFT}
\label{Sec:JTgravity}

 This section is devoted to the review on the essential ingredients needed for our discussion.
 We first address the features of the Jackiw-Teitelboim (JT) gravity,  focusing on the dS$_2$ background.
When the JT gravity is coupled to CFT$_2$, the entanglement entropy of radiation can be produced, the expression of which will be presented as well.

 \subsection{JT gravity on dS$_2$}
 
 The JT gravity is a theory of   two-dimensional gravity coupled to dilaton, which can be obtained by the dimensional reduction of   three-dimensional gravity.
 For the dS$_3$ background, the action is written as 
  \dis{S=\frac{1}{16\pi G^{(3)}}\int d^3 x \sqrt{-g^{(3)}}\Big(R^{(3)}-2 H^2\Big)-\frac{1}{8\pi G^{(3)}}\int d^2x \sqrt{-h^{(3)}}K^{(3)}.}
  Under the ansatz for the dimensional reduction,
  \dis{ds_3^2=g^{(2)}_{ij}dx^i dx^j+\Big(\frac{\Phi(x^i)}{2\pi H}\Big)^2d\varphi^2,\label{eq:metric3to2}}
 we obtain
  \dis{&R^{(3)}=R^{(2)}-\frac{2}{\Phi}\square^{(2)}\Phi,
  \\
  &K^{(3)}=K^{(2)}+{g^{(3)}}^{\varphi\varphi}K_{\varphi\varphi}=K^{(2)}+\frac{1}{\Phi}n^\mu \nabla^{(2)}_\mu \Phi, \label{eq:curvatures}}
  where $n^\mu$ is the unit vector normal to the boundary surface, from which the action is  reduced to that of the JT gravity,
 \dis{S=\int d^2x \sqrt{-g^{(2)}}\frac{\Phi}{16\pi G}\Big(R^{(2)}-2 H^2\Big)-\int d x\sqrt{-h^{(2)}}\frac{\Phi_b}{8\pi G} K^{(2)}.\label{eq:JTaction}}
 Here the two-dimensional Newton's constant is given by $G=G^{(3)} H$.
 Since the JT gravity and dS$_3$ are connected through the dimensional reduction, the solutions $(g^{(2)}_{ij}, \Phi)$ to the Einstein equation can be immediately read off from the components of dS$_3$ metric.
Here we list the solutions in various  coordinates for dS space.

 \begin{figure}[!t]
  \begin{center}
   \includegraphics[width=0.5\textwidth]{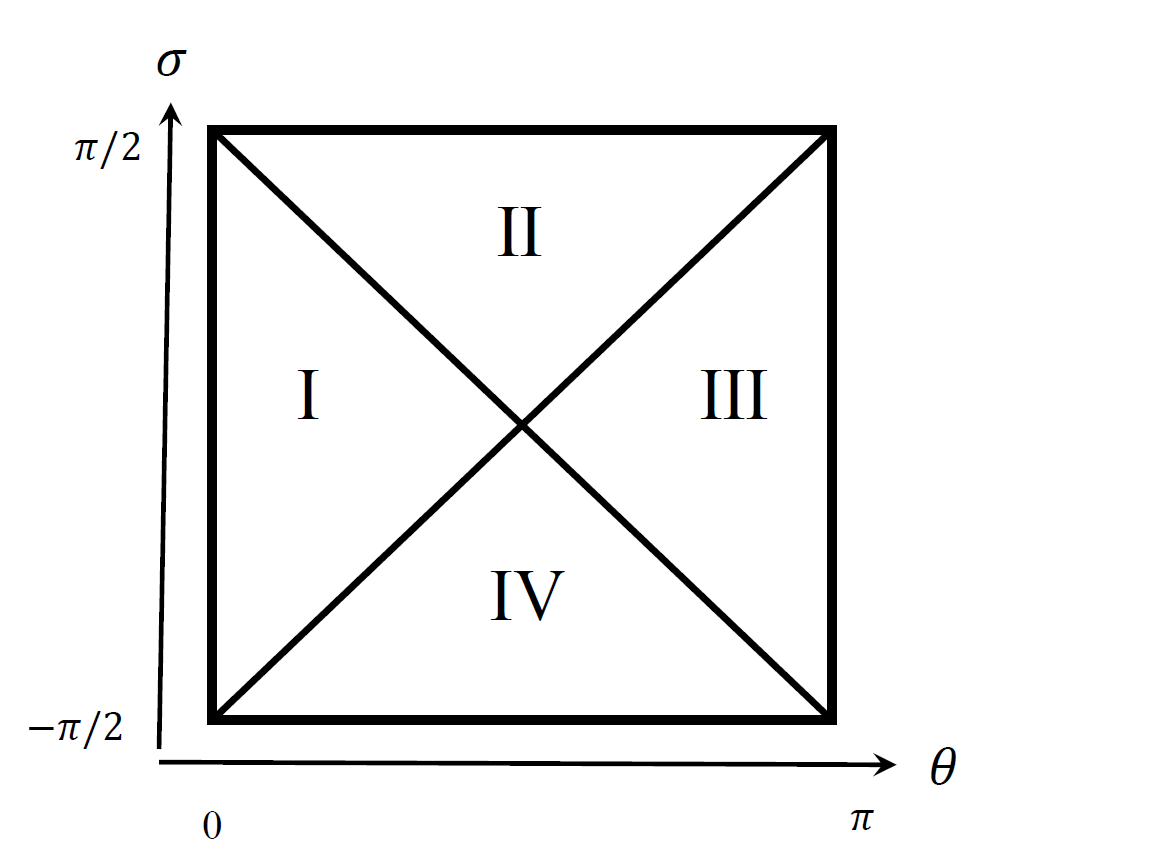}
  \end{center}
 \caption{Penrose diagram for dS$_2$. 
 The conformal coordinates $(\sigma, \theta)$ cover the whole region, regions I, II, III, and IV.
  }
\label{fig:global}
\end{figure}

\begin{itemize}
\item Conformal coordinates $(\sigma, \theta)$ :  The metric and dilaton are given by
\dis{ds_2^2=\frac{1}{H^2 \cos^2\sigma }(-d\sigma^2+d\theta^2),\quad\quad \Phi=2\pi\frac{\sin\theta}{\cos\sigma}.}
Here $\sigma \in (-\pi/2, \pi/2)$ and $\theta \in (0, \pi)$, which  cover the whole dS$_2$ manifold (regions I, II, III, and IV) as depicted in Fig. \ref{fig:global}.
When we describe dS$_2$  as the hyperboloid embedded in 3-dimensional Minkowski space satisfying $-(X^0)^2+(X^1)^2+(X^2)^2=H^{-2}$, the conformal coordinates correspond to the parametrization
\dis{X^0=H^{-1}\tan\sigma, \quad X^1=H^{-1}\frac{\sin\theta}{\cos\sigma},\quad X^2=H^{-1}\frac{\cos\theta}{\cos\sigma}.\label{eq:global}}

 \begin{figure}[!t]
  \begin{center}
   \includegraphics[width=0.4\textwidth]{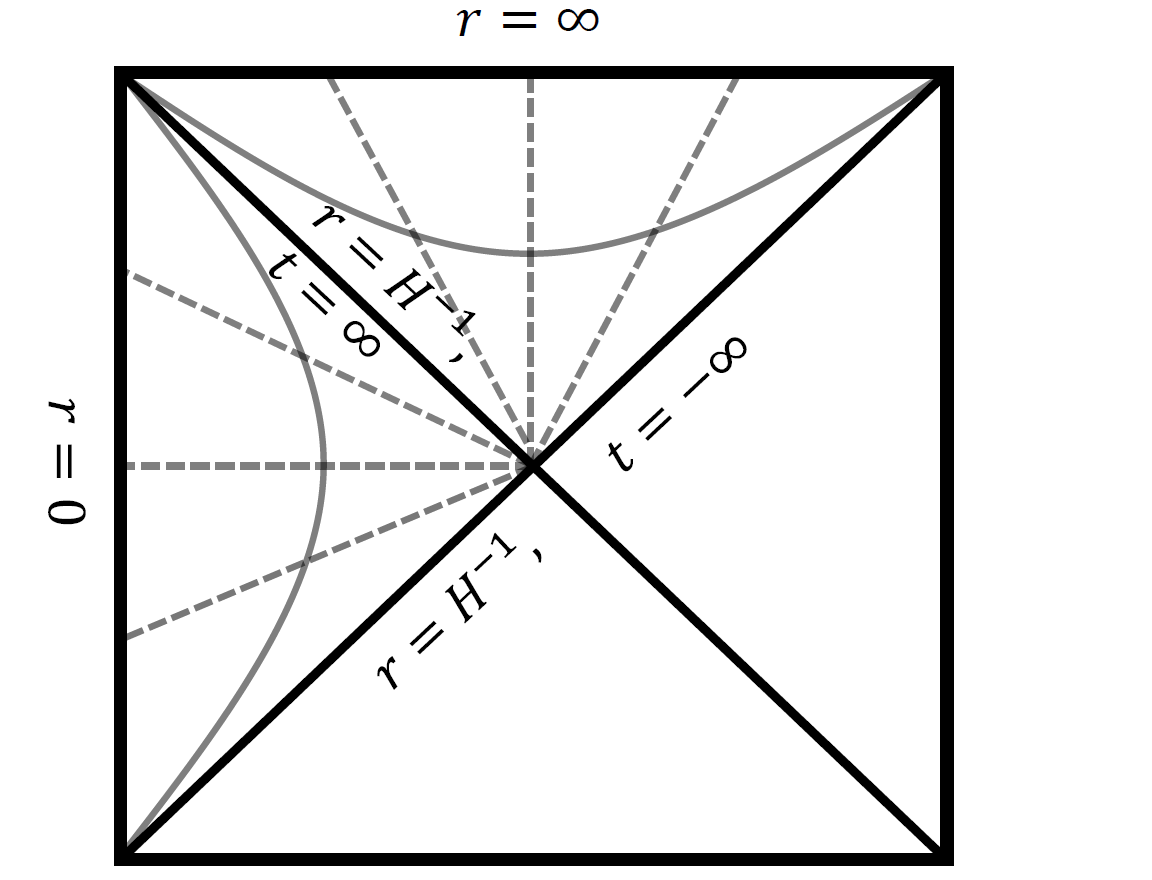}
   \includegraphics[width=0.4\textwidth]{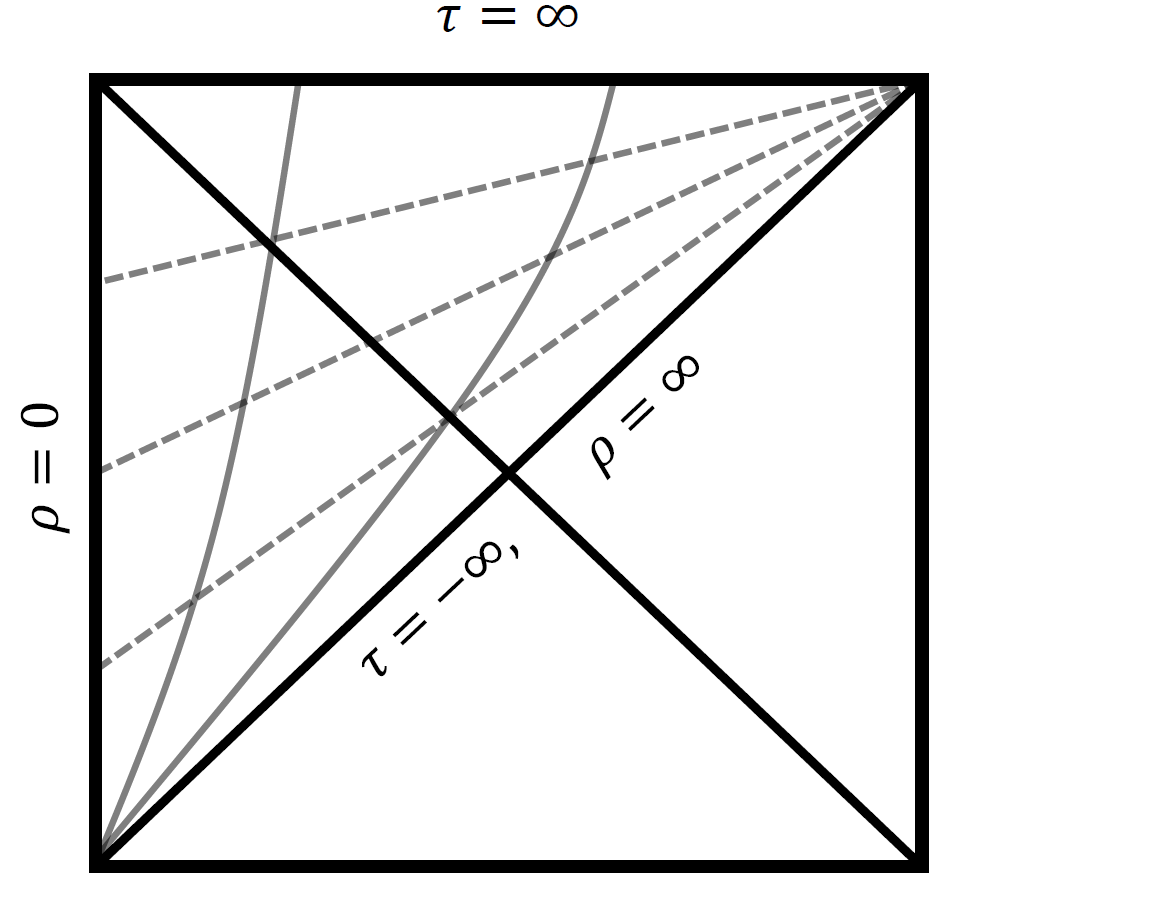}
  \end{center}
 \caption{(Left) :  Covering of  regions I (static patch) and II by their own static coordinates. 
 Here $t$ ($r$) is constant on grey dashed (grey) lines.
(Right) : Covering of the inflating patch, the union of regions I and II by the flat coordinates. 
 Here $\tau$ ($\rho$) is constant on grey dashed (grey) lines.
  }
\label{fig:dScoord}
\end{figure}

\item Static coordinates $(t, r)$ : The metric and dilaton are given by
\dis{ds_2^2=-(1-H^2r^2)dt^2+\frac{dr^2}{1-H^2 r^2},\quad\quad \Phi= 2\pi H r.}
For the parametrization giving above solutions,
\dis{X^0=H^{-1}\sqrt{1-H^2r^2}\sinh(Ht),\quad X^1=r,\quad X^2=H^{-1}\sqrt{1-H^2r^2}\cosh(Ht),\label{eq:staticI}}
the coordinates  $t\in (-\infty, +\infty)$ and $r \in (0, H^{-1})$ cover the static patch (region I in Fig. \ref{fig:global}).
Meanwhile,  region II can be covered by another parametrization 
\dis{X^0=H^{-1}\sqrt{H^2r^2-1}\cosh(Ht),\quad X^1=r,\quad X^2=H^{-1}\sqrt{H^2r^2-1}\sinh(Ht),\label{eq:staticII}}
 as $t \in (-\infty, +\infty)$ and $r\in (H^{-1}, \infty)$.
 The covering of  regions III and IV can be obtained by the replacement  of  $(X^0, X^1, X^2)$ in \eqref{eq:staticI} and \eqref{eq:staticII} by $(-X^0, X^1, -X^2)$, respectively,  which corresponds to $\sigma \to -\sigma$ and  $\theta \to \pi-\theta$ in terms of \eqref{eq:global}.
In any case, the static coordinates are regular only in the region they are defined and different regions are separated by the coordinate singularity at the horizon, $r=H^{-1}$.
The covering of region I and II by the static coordinates   is depicted in the left panel of Fig. \ref{fig:dScoord}.

\item Flat coordinates $(\tau, \rho)$ : The metric and dilaton are given by
\dis{ds_2^2=-d\tau^2+e^{2 H\tau} d\rho^2, \quad\quad \Phi=2\pi H\rho e^{H\tau},}
where $\tau \in (-\infty, +\infty)$ and $\rho \in (0, \infty)$.
This corresponds to the parametrization
\dis{X^0=H^{-1}\Big(\sinh(H\tau)+\frac12(H\rho)^2e^{H\tau}\Big), X^1=\rho e^{H\tau}, X^2=H^{-1}\Big(\cosh(H\tau)-\frac12(H\rho)^2e^{H\tau}\Big).}
 The static and flat coordinates are related by
 \dis{t=\tau-\frac{1}{2H}\log\Big(1-H^2\rho^2 e^{2H\tau}\Big),\quad r=\rho e^{H\tau}\label{eq:stflI}}
 in region I and
 \dis{t=\tau-\frac{1}{2H}\log\Big(H^2\rho^2 e^{2H\tau}-1\Big),\quad r=\rho e^{H\tau}}
 in region II, respectively.
 The flat coordinates cover the inflating patch, the union of regions I and II, without the coordinate singularity, as can be seen in the right panel of Fig. \ref{fig:dScoord}.
\end{itemize}

 In the following discussions, we consider $H$ varying with respect to some  timelike direction which we will call $\tau$.
 In this case, the hypersurface orthogonal to the $\tau$ direction has a rotational invariance as an isometry.
 Then we expect  the corrected metric to be in the form of FRW metric, in which $e^{H\tau}$ in \eqref{eq:stflI} is replaced by the scale factor $a(\tau)$ with $a^{-1}da/d\tau$ is close to the constant $H$.
 Indeed, the dimensional reduction  shows that, up to surface term which does not affect the equations of motion, the action for a $\tau$ dependent $H$ can be obtained by replacing the constant $H$ by $H(\tau)$.
 We sketch the derivation of this in App. \ref{app:modS}.

 We remark that the value of dilaton $\Phi$ is positive in any coordinates and  interpreted as an area of surface in dS$_3$   on which the dS$_2$ coordinates  (time and radial coordinates) are fixed.
We also note that   two-dimensional gravity allows the Gauss-Bonnet term, which shifts   $\Phi$ by some constant $\Phi_0$.
This indeed what happens when we couple the JT gravity  to CFT$_2$, which we will study in detail throughout this article.
To see this, we recall that the trace anomaly provides (see, e.g., \cite{Aguilar-Gutierrez:2021bns})
\dis{\int d^2 x\sqrt{-g^{(2)}}\langle T^\mu_\mu\rangle=\frac{c}{24\pi}\int d^2 x \sqrt{-g^{(2)}} R^{(2)},\label{eq:shiftD}}
which shifts $\Phi$ by $(2/3)c G=(2/3)c G^{(3)}H$.
Thus, the area term in the entropy contains the central charge dependent term as well.


\subsection{CFT entropy in dS$_2$ background}

 We now consider  CFT$_2$ coupled to the JT gravity.
 In terms of the lightcone coordinates
\dis{x^+=e^{i (\theta-\sigma)},\quad\quad x^-=e^{-i (\theta+\sigma)},}
the dS$_2$ metric is written in the form of the Weyl rescaling of the flat metric,
\dis{ds^2=\frac{dx^+ dx^-}{\Omega(x^+, x^-)^2} ,\quad\quad  \Omega(x^+, x^-)=\frac{H}{2}(1+x^+x^-).}
Then the entanglement entropy, or the von Neumann entropy of the CFT matter on a segment between $(x_1^+, x_1^-)$ and $(x_2^+, x_2^-)$  in dS$_2$ can be written as 
\dis{S_{\rm mat}&=\frac{c}{6}\log\Big[\frac{1}{\epsilon_{\rm UV}^2}\Big(\frac{(x^+_1-x^+_2)(x^-_1-x^-_2)}{\Omega(x_1)\Omega(x_2)}\Big)\Big]
\\
&=\frac{c}{6}\log\Big[\frac{2}{\epsilon_{\rm UV}^2 H^2}\Big(\frac{\cos(\sigma_1-\sigma_2)-\cos(\theta_1-\theta_2)}{\cos\sigma_1\cos\sigma_2}\Big)\Big],}
where $c$ is the central charge of CFT$_2$ and $\epsilon_{\rm UV}$ is the UV cutoff length  \cite{Holzhey:1994we, Calabrese:2004eu, Calabrese:2009qy}.
While the expression above is given in terms of the conformal coordinates, it is straightforward to rewrite this in terms of other coordinates  by noting that the argument of the logarithm  is nothing more than the spacetime interval   in three-dimensional Minkowski space $(X_1 - X_2)^2$, i.e.,
\dis{&\frac{\cos(\sigma_1-\sigma_2)-\cos(\theta_1-\theta_2)}{\cos\sigma_1\cos\sigma_2} = 1+\tan\sigma_1 \tan\sigma_2-\frac{\sin\theta_1}{\cos\sigma_1}\frac{\sin\theta_2}{\cos\sigma_2}-\frac{\cos\theta_1}{\cos\sigma_1}\frac{\cos\theta_2}{\cos\sigma_2}
\\
&=1+H^2(X_1^0 X_2^0-X_1^1 X_2^1-X_1^2 X_2^2)=1-H^2 X_1\cdot X_2 =\frac{H^2}{2}(X_1-X_2)^2,}
such that
\dis{S_{\rm mat}=\frac{c}{6}\log\Big[\frac{(X_1-X_2)^2}{\epsilon_{\rm UV}^2}\Big].}
In the flat coordinates, it can be written as
\dis{S_{\rm mat}=\frac{c}{6}\log\Big[\frac{2}{\epsilon_{\rm UV}^2 H^2}\Big(1-\cosh[H(\tau_1-\tau_2)]+\frac12 e^{H(\tau_1+\tau_2)}H^2(\rho_1-\rho_2)^2\Big)\Big].}
In the static coordinates, it is written as
\dis{S_{\rm mat}=\frac{c}{6}\log\Big[\frac{2}{\epsilon_{\rm UV}^2 H^2}\Big(1- H^2 r_1r_2-\sqrt{1-H^2r_1^2}\sqrt{1-H^2r_2^2}\cosh[H(t_{1}-t_{2})]\Big)\Big]\label{eq:entstat}}
for region I and
\dis{S_{\rm mat}=\frac{c}{6}\log\Big[\frac{2}{\epsilon_{\rm UV}^2 H^2}\Big(1- H^2 r_1r_2+\sqrt{H^2r_1^2-1}\sqrt{H^2r_2^2-1}\cosh[H(t_{1}-t_{2})\Big)\Big]}
for region II, respectively.

\section{Information paradox in quasi-dS static patch}
\label{Sec:dSparadox}

\subsection{Choice of spacelike surface collecting radiation}
\label{Sec:Sigma}

 Since a static observer in (quasi-)dS space detects the thermal   radiation  emitted by the horizon,  the observer's  semiclassical description of (quasi-)dS space is expected to be quite similar to that of black hole as seen from far outside the horizon.
 This may motivate one to postulate the central dogma for (quasi-)dS space as follows:
\begin{quote} 
 (quasi-)dS space as seen by an static observer surrounded by the horizon is described by the unitarily evolving quantum system with exp$[({\rm Horizon~Area})/(4G)]$ degrees of freedom.
 \end{quote}
  This slight modification of the central dogma for black hole, however, conceals a number of subtleties. 
 For black hole, a compact object in space, it is natural to introduce the cutoff surface outside which gravitational effects are small enough to be neglected.
 Then  the region inside the cutoff surface is treated as the quantum black hole, the quantum gravity system  in the central dogma. 
 The radiation entropy is measured by collecting the radiation passing through the spacelike surface extending over the region outside the cutoff surface.
 In contrast, for  (quasi-)dS space, the curvature does not vanish at any point  :  the Ricci scalar of perfect dS$_d$ is given by a constant, $R^{(d)}=d(d-1)H^2$.
 Thus, gravity cannot be neglected at any point in (quasi-)dS space and the separation of the observer collecting the radiation from the quantum gravity system is challenging.
 Nevertheless, when $H (G^{(d)})^{1/(d-2)} \ll 1$, the most part of the static patch can be treated semiclassically, in which the curved spacetime is just a non-dynamical background.
 \footnote{Even though this condition does not apply to the case of $d=2$, the two dimensional gravity we are discussing is a dimensional reduction of three dimensional gravity, thus the condition $HG^{(3)} \ll 1$ is assumed.}
 Then we can consider the spacelike surface $\Sigma$ in this `radiation region' (or thermal bath) to collect the radiation.
 At the same time, the union of the rest part of the static patch and the region beyond the horizon is regarded as the quantum gravity system  in the central dogma.

 Before considering quasi-dS space, we address the radiation entropy   in perfect dS space in which $H$ is constant.
If we choose the spacelike surface $\Sigma$ to be the equal-time surface in the conformal or flat coordinates,  $\sigma=$(constant) or $\tau=$(constant), the entanglement entropy of the CFT radiation  given by
 \dis{S_{\rm mat}=\frac{c}{6}\log\Big[\frac{2}{\epsilon_{\rm UV}^2 H^2}\Big(\frac{1-\cos(\theta_1-\theta_2)}{\cos^2\sigma}\Big)\Big]\quad\quad &\sigma=({\rm constant}),\quad{\rm or}
 \\
S_{\rm mat}=\frac{c}{6}\log\Big[\frac{1}{\epsilon_{\rm UV}^2  } e^{2H \tau } (\rho_1-\rho_2)^2 \Big] \quad\quad &\tau=({\rm constant}),\label{eq:flatentropy}}
where $\theta_{1,2}$ and $\rho_{1,2}$ being the coordinates of two endpoints of $\Sigma$ in each case, explicitly depends on the time coordinate $\sigma$ or $\tau$.
 Let us focus on the flat coordinates which are appropriate to describe the homogeneous and isotropic universe.
 They are also useful if we are interested in the spacelike future infinity, which is taken as the boundary of dS space in the dS/CFT correspondence \cite{Strominger:2001pn}.
We infer from \eqref{eq:flatentropy} that in the flat coordinates, $S_{\rm mat}$ increases in $\tau$ for fixed $\rho_1-\rho_2$.
Then one is tempted to argue that even in perfect dS space, we can find the direction of irreversibility  parametrized by $\tau$ (for previous works along the similar line, see, e.g., \cite{Hu:1986jd, Brandenberger:1992sr, Brandenberger:1992jh, Prokopec:1992ia, Gasperini:1992xv, Gasperini:1993mq, Brahma:2020zpk}).
However, as often summarized as ``perfect dS space does not have a clock", the geometries at $\tau$ and $\tau+a$ are not distinct.
This is because the shift in time $\tau \to \tau+a$ can be compensated by the rescaling  $\rho \to e^{-a H} \rho$ to leave the geometry unchanged, which indeed is one of dS isometries.
  Since the static radial coordinate given by $r=\rho e^{H\tau}$ is invariant under this transformation, the absence of the clock is obvious if we choose $\Sigma$ to be a surface $t=$(constant) where $t$ is the static time coordinate, with the $r$ coordinates of the endpoints of $\Sigma$ fixed.
  Then we expect that $S_{\rm mat}$ on this surface does not evolve in time. 
    This is consistent with the observation that in the Bunch-Davies vacuum in which dS isometries are preserved, the perfect dS space is in thermal equilibrium so the thermodynamic quantities like the horizon size are static.
The clock can be introduced in quasi-dS space in which $H$ varies in time, as the dS isometry associated with the shift in $\tau$ and the rescaling of $\rho$ is spontaneously broken.
In this case, $S_{\rm mat}$ on the surface $t=$(constant)  will evolve in time, measuring the irreversibility generated by the out-of-equilibrium process analogous to the  black hole evaporation.
  Moreover, in the static coordinates, unlike the flat coordinates, regions I (inside the horizon) and II (beyond the horizon) are separated by the coordinate singularity hence the role of the horizon is emphasized.
  This indicates that the static coordinates enable one to investigate thermodynamics relevant to the horizon more manifestly from the direct comparison with the black hole thermodynamics.

 \begin{figure}[!t]
  \begin{center}
   \includegraphics[width=0.5\textwidth]{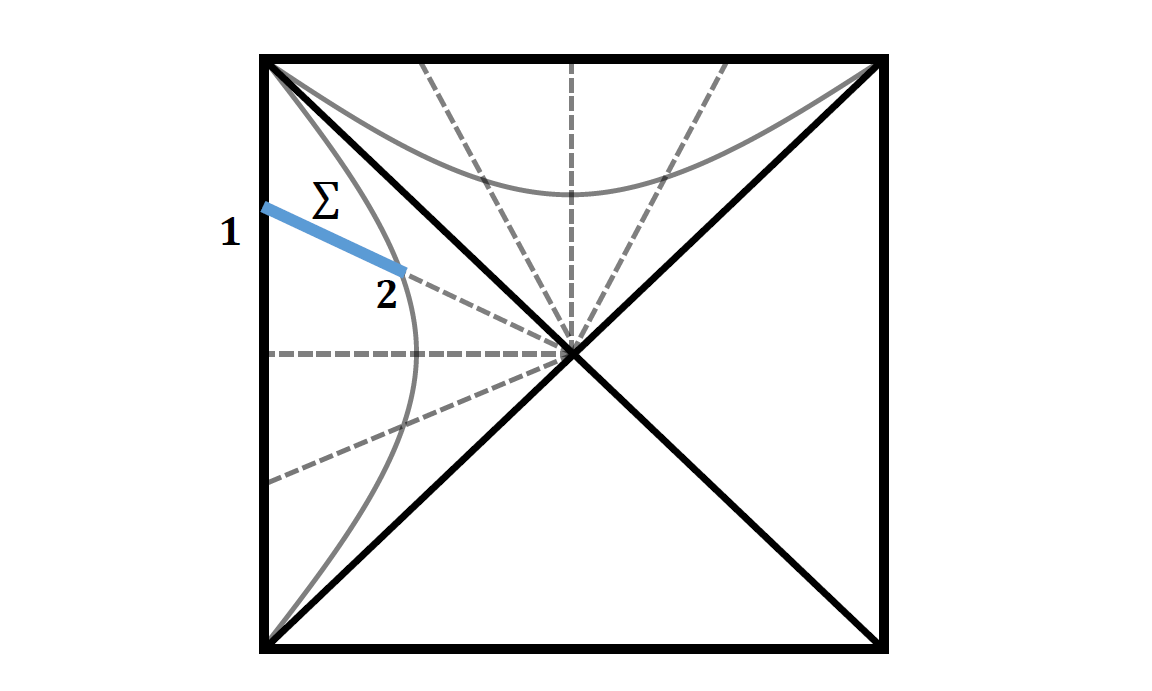}
  \end{center}
 \caption{The thick blue line indicates our choice of $\Sigma$ lying on the $t=$(constant) surface to collect the radiation.
 The static coordinates of the endpoints $1$ and $2$ are given by $(t_1, r_1)=(t, 0)$ and $(t_2, r_2)=(t, r)$, respectively.
  }
\label{fig:Radsurf}
\end{figure}

  From the discussion above, we take $\Sigma$ to lie on the spacelike surface $t=$(constant) within the static patch  such that the static coordinates of two endpoints  are given by  $(t_1, r_1)=(t, 0)$ and $(t_2, r_2)=(t, r)$ ($r< H^{-1}$), respectively,  as shown in Fig. \ref{fig:Radsurf}.  
  Then for perfect dS space,  the radiation entropy   on $\Sigma$ is written as
  \dis{S_{\rm mat}(\Sigma)= \frac{c}{6}\log\Big[\frac{2}{\epsilon_{\rm UV}^2 H^2}\Big(1-   \sqrt{1-H^2r^2} \Big)\Big].\label{eq:SraddS}}
 This explicitly shows that  $S_{\rm mat}(\Sigma)$ does not evolve  in time $t$, reflecting the thermal equilibrium under the dS isometries.
 While the central charge is typically taken to be $c \gg 1$ in order to suppress the subleading corrections which are not taken into account in our discussion, it cannot be arbitrarily large since $S_{\rm mat}(\Sigma)$ is restricted to be smaller than the dS entropy.
 Taking the shift of the dilaton $\Phi \to \Phi+(2/3)cG$  coming from \eqref{eq:shiftD}  into account, the dS entropy is given by
 \dis{S_{\rm dS}=\frac{\rm Horizon~Area}{4G} = \frac{\pi}{2 G}+\frac{c}{6}=\frac{\pi}{2G^{(3)} H}+\frac{c}{6},\label{eq:dSent}}
 where  the horizon area is given by $\Phi(r)=2\pi H r$ with $r=H^{-1}$.
  
 We note that in the JT gravity, the dilaton multiplied to the Ricci tensor in the action \eqref{eq:JTaction} seems to enhance the strength of the gravitational interaction at $r=0$, the endpoint of $\Sigma$, as the Newton's constant reads $G/\Phi=G^{(3)}/(2\pi r)$ effectively.
 This may invalidate our setup that gravity is negligible over the region of $\Sigma$.
 When we regard the JT gravity as  a dimensional reduction of  three dimensional gravity, such an enhancement is an artifact caused by promoting the factor $r$ in $(-g^{(3)})^{1/2}$ which was originally the component of the three dimensional metric in the angular direction to the dilaton field $\Phi$.
 Nevertheless, one may circumvent the issue of strong gravity by  taking one of endpoints of $\Sigma$ to be $(t_1, r_1)=(t, \epsilon_{\rm UV})$  and imposing $G^{(3)}\ll \epsilon_{\rm UV}$.
 So fas as $\epsilon_{\rm UV}$ is much smaller than $H^{-1}$, we can neglect the $\epsilon_{\rm UV}$ dependence in $S_{\rm mat}(\Sigma)$.

 \subsection{Radiation entropy in quasi-dS space}
 
  In quasi-dS space, $H$ is no longer constant in time as the dS isometries are spontaneously broken.
  The radiation  in this case is not in equilibrium with the quantum gravity system, which leads to the evolution of  the radiation entropy  in time.
  In order to see the implications to the inflationary cosmology, we consider $H=H(\tau)$ as a monotonically decreasing function of the flat time coordinate $\tau$.
  Then the deviation of the background from  perfect dS space can be treated in a controllable way through the expansion in terms of slow-roll parameters, 
      \dis{\epsilon_H = -\frac{1}{H^2}\frac{dH}{d\tau},\quad\quad
 \eta_H=-\frac{1}{2H}\frac{d^2H/d\tau^2}{dH/d\tau},}
 where $\epsilon_H >0$.

   Meanwhile,  as discussed in Sec. \ref{Sec:Sigma}, the surface $\Sigma$ is taken to be the equal-static-time ($t$) surface.
   It is reasonable to assume here that  the leading term of $S_{\rm mat}(\Sigma)$ for the quasi-dS background is given by $(c/6)\log[(X_1-X_2)^2/\epsilon_{\rm UV}^2]$, the same form as $S_{\rm mat}(\Sigma)$ for the perfect dS background. 
  Then different $H$ values at two endpoints  of $\Sigma$ make $S_{\rm mat}(\Sigma)$ time dependent.
  More explicitly, from the relation between $t$ and $\tau$ in \eqref{eq:stflI},  the flat time coordinates of $X_1$ and $X_2$ are given by
   \dis{\tau_1=t,\quad\quad \tau_2=t+\frac{1}{2H_2}\log(1-H_2^2r^2),}
  respectively, where $H_{1,2}=H(\tau_{1,2})$.   
  Since  $1-H_2^2r^2 <1$ one finds that $\tau_1 >\tau_2$ hence $H_1 < H_2$.  
 For $\epsilon_H \ll 1$, the values of $H_1$ and $H_2$ are similar in magnitude.
 Then we can expand $H_2$ as
\dis{H_2=H_1\Big[1  -\epsilon_H (t) \Big(\frac{H_1 }{2H_2}\log(1-H_2^2 r^2)\Big)+ \epsilon_H \eta_H \Big(\frac{H_1 }{2H_2}\log(1-H_2^2 r^2)\Big)^2+\cdots\Big].\label{eq:H1H2}}
Given $\eta_H \simeq \epsilon_H\ll 1$, we focus on the value of $r$ satisfying
\dis{\epsilon_H (t) \Big[\frac{H_1 }{2H_2}\log\Big(\frac{1}{1-H_2^2 r^2}\Big)\Big] < 1,}
for the controllable  expansion in terms of slow-roll parameters.
While this obviously requires $\epsilon_H |\log(1-H_1^2r^2)| <1$, the expansion of LHS,
\dis{\frac{\epsilon_H}{2}\log(1-H_1^2 r^2)+\frac{\epsilon_H^2}{2}\log(1-H_1^2 r^2)\Big[\frac{1}{2}\log(1-H_1^2 r^2)+\frac{H_1^2r^2}{1-H_1^2 r^2}\Big]+{\cal O}(\epsilon_H^3),}
indeed imposes more stringent constraint on $r$,
\dis{\frac{\epsilon_H H_1^2 r^2}{1-H_1^2r^2} <1,\label{eq:rconst}}
preventing $r$ from being too close to the horizon $H^{-1}$.

 When the radiation region occupies the most part of the static patch,  $r$ is taken to be close to the horizon so far as \eqref{eq:rconst} is satisfied.
 In this case, it is convenient to express $S_{\rm mat}(\Sigma)$ in terms of  $\ell$ defined as
\dis{H_2^2 \ell^2 = 1-H_2^2 r^2,\label{eq:ell}}
which is much less than $1$,  measuring the deviation of the endpoint $X_2$ from the horizon. 
Then the condition \eqref{eq:rconst} reads $\epsilon_H < (H_1 \ell)^2$ (since $H_1 \simeq H_2$ and $H_2\ell\ll 1$) and \eqref{eq:H1H2} is rewritten as
\dis{H_1-H_2=\epsilon_H \frac{H_1^2}{2H_2}\log ( H_2^2\ell^2 )+{\cal O}(\epsilon_H^2) =-\epsilon_H H_1\log\Big(\frac{1}{H_2\ell}\Big)+{\cal O}(\epsilon_H^2).}
Therefore, given the radiation entropy,
\dis{S_{\rm mat}(\Sigma)&=\frac{c}{6}\log\Big[\frac{(X_1-X_2)^2}{\epsilon_{\rm UV}^2}\Big]
\\
&=\frac{c}{6}\log\Big[\frac{1}{\epsilon_{\rm UV}^2 H_1H_2}\Big(\frac{H_2}{H_1}+\frac{H_1}{H_2}-2 \sqrt{1-H_2^2r^2}\cosh[(H_1-H_2)t]\Big)\Big],}
 the leading term  in the expansion with respect to $\epsilon_H$ is written as
 \dis{S_{\rm mat}(\Sigma)=\frac{c}{6}\log\Big[\frac{2}{\epsilon_{\rm UV}^2 H_1H_2}\Big(1- H_2\ell\cosh\Big[\epsilon_H H_1t \log\Big(\frac{1}{H_2\ell}\Big)\Big]\Big)\Big].\label{eq:RadEnt}}
We will observe   the time evolution of  \eqref{eq:RadEnt}  with  $\ell$ fixed.
 We note that the expression \eqref{eq:RadEnt} is not valid for arbitrary large $t$.
 For $ H_1 t \gg 1$, the argument of hyperbolic cosine becomes larger than $1$ such that the term in the parentheses is approximated as

 \dis{1-H_2 \ell \cosh\Big[\epsilon_H H_1t \log\Big(\frac{1}{H_2\ell}\Big)\Big] \simeq 1-\frac12 (H_2\ell )^{1-\epsilon_H H_1 t}. }
 Since this  term is required to be positive, 
   \footnote{More precisely, the term $1-H_2\ell \cosh[\epsilon_H H_1 t \log[1/(H_2\ell)]]$ in the logarithm of \eqref{eq:RadEnt} is bounded by $\epsilon_{\rm UV}^2 H_1 H_2$ as $(X_1-X_2)^2 \geq \epsilon_{\rm UV}^2$.
  But since  $\epsilon_{\rm UV}^2 H_1 H_2 \ll 1$, our estimation is not much affected.}  
   we find a bound $t < (\epsilon_H H_1)^{-1}(\log[H_1\ell/2]/\log[H_1\ell])$, or roughly $t < (\epsilon_H H_1)^{-1}$ up to ${\cal O}(1)$ coefficient.
 In fact, the bound on $t$ we found  is nothing more than the time scale after which $H_2$ is no longer close to $H_1$.
 One way to see this is to observe the explicit form of $H(\tau)$ for the special case of a constant $\epsilon_H$, $H(\tau)=H_0/(1+\epsilon_H H_0\tau)$, which significantly deviates from $H_0$ after $\tau>(\epsilon_H H_0)^{-1}$.
 
 Finally, the central charge $c$ which is proportional to the number of degrees of freedom may vary in time. 
 Here we will focus on the case that $c$ is monotonically increasing in time.
 This is motivated by the entropy argument for the dS swampland conjecture \cite{Ooguri:2018wrx}, which is based on the claim that the number of low energy degrees of freedom   increases exponentially along the trans-Planckian geodesic trajectory of the modulus responsible for the vacuum energy \cite{Ooguri:2006in}.
   In our analysis,  for simplicity, we assume that in region I, $c$ depends only on the static time coordinate $t$ such that it is constant on the slice $\Sigma$.
 In  fact, for the application to the inflationary cosmology in which   homogeneity and isotropy are emphasized, it is reasonable to take $c$ to be a function of the flat time coordinate $\tau$, just like $H$.
  To compare the difference between the values of $c$ at two endpoints of $\Sigma$ in this case, we employ the ansatz motivated by the distance conjecture \cite{Ooguri:2006in}, $c = c_0 {\rm exp}[\lambda \Delta \phi]$,  where $\Delta \phi$ is the geodesic distance traversed by the modulus in Planck unit and $\lambda$ is an ${\cal O}(1)$ constant.
  In the four-dimensional case, the equations of motion are solved to give $d\phi/d\tau = \sqrt{2\epsilon_H} H$, implying $dc/d\tau = \lambda \sqrt{2\epsilon_H} H c $.
  This suggests to consider $c(\tau)$ satisfying $[c(\tau_1)-c(\tau_2)]/c(\tau_1) ={\cal O}(\epsilon_H^{1/2})$, in which case $S_{\rm mat}(\Sigma)$ containing $c(t)$ and $c(\tau)$  differ by the ${\cal O}(\epsilon_H^{1/2})$ correction.
  While this dominates over the ${\cal O}(\epsilon_H)$ correction coming from difference between $H_1$ and $H_2$ discussed above, it does not change the qualitative feature of the conclusion we will draw so we will restrict our attention to $c(t)$ rather than $c(\tau)$.

 \subsection{Condition for information paradox} 
  
 Now we consider  the condition for the information paradox in three-dimensional quasi-dS space,  regarding the JT gravity as a dimensional reduction of dS$_3$ space.
 For this purpose, we set $G^{(3)}$ rather than $G= G^{(3)} H$ to be a constant.
 For the information paradox to arise in quasi-dS space, the increasing rate of $S_{\rm mat}(\Sigma)$ in time must be larger than that of $S_{\rm dS}$ such that $S_{\rm mat}(\Sigma)$ eventually exceeds $S_{\rm dS}$.
Since  $H$ depends only on $\tau$, not on $\rho$, we have $dH/dt=dH/d\tau=-\epsilon_H H^2$,
\footnote{ Derivative with respect to the static time coordinate $t$ is written in terms of derivatives with respect to  the flat coordinates $(\tau, \rho)$ as
\dis{\frac{d}{dt}=\frac{\partial \tau}{\partial t}\frac{\partial}{\partial \tau}+\frac{\partial \rho}{\partial t}\frac{\partial}{\partial \rho}=\frac{\partial}{\partial \tau}- H\rho\frac{\partial}{\partial \rho}.}
}
 from which one finds that the derivative of $S_{\rm dS}$ given by \eqref{eq:dSent} with respect to $t$ becomes
\dis{\frac{d S_{\rm dS}}{dt}=-\epsilon_H H_1^2 \frac{d S_{\rm dS}}{dH_1}=\frac{\pi \epsilon_H}{2G^{(3)}} +\frac16\frac{dc}{dt} >0.}
This shows that $S_{\rm dS}$ increases in time provided $dc/dt$ is positive or suppressed to ${\cal O}(\epsilon_H^2)$.
The same feature is found in the four-dimensional case, in which $S_{\rm dS}=\pi/(GH^2)$, giving $dS_{\rm dS}/dt=2\pi \epsilon_H/(GH) > 0$.

 On the other hand, the radiation entropy $S_{\rm mat}(\Sigma)$ given by \eqref{eq:RadEnt} decreases in time as $t$ approaches the bound $(\epsilon_H H_1)^{-1}$, since  the argument of the logarithm vanishes for $t\sim (\epsilon_H H_1)^{-1}$.
  However, $S_{\rm mat}(\Sigma)$  also has an implicit $t$ dependence through $H_1$ and $c$, which  provides a chance for $S_{\rm mat}(\Sigma)$ to increase in $t$ and exceed $S_{\rm dS}$ during some period.
 Then the information paradox arises in quasi-dS space.
 To see this more explicitly, we take  the derivative of $S_{\rm mat}(\Sigma)$ with respect to $t$ using
 \dis{\frac{d}{dt}=\frac{\partial}{\partial t} - \epsilon_H H_1^2 \frac{\partial}{\partial H_1}+\frac{dc}{dt}\frac{\partial}{\partial c}.} 
To leading order in $\epsilon_H$, $dS_{\rm mat}(\Sigma)/dt$ can be written as a sum of four terms, $dS_{\rm mat}(\Sigma)/dt = \dot{S}_c + \dot{S}_{H1} +\dot{S}_t +\dot{S}_{H2}$:
\begin{itemize}
\item $\dot{S}_c$ : This term comes from the time evolution of $c$.
\dis{\dot{S}_c = \frac{dc/dt}{c}S_{\rm mat}(\Sigma)=\frac{dc/dt}{6}\log\Big[\frac{2}{\epsilon_{\rm UV}^2 H_1H_2}\Big(1- H_2\ell \cosh\Big[\epsilon_H H_1t \log\Big(\frac{1}{H_2\ell}\Big)\Big]\Big)\Big].\label{eq:Sdc}}
\item $\dot{S}_{H1}$ : This term comes from the time evolution of $H_1$ in  $(c/6)\log[2/(\epsilon_{\rm UV}^2 H_1 H_2)]$ part of \eqref{eq:RadEnt}.
\dis{\dot{S}_{H1}=\frac{c}{3}\epsilon_H H_1.\label{eq:SdH1}}
\item $\dot{S}_t$ : This term comes from the explicit $t$ dependence in $(c/6)\log[X(t)]$ part of \eqref{eq:RadEnt}, where
\dis{X(t)= 1- H_2\ell \cosh\Big[\epsilon_H H_1t \log\Big(\frac{1}{H_2\ell}\Big)\Big] .}
\dis{\dot{S}_t = -\frac{c}{6}\epsilon_H H_1\Big( H_1\ell \log\Big[\frac{1}{H_1\ell}\Big]\Big)\frac{\sinh\Big[\epsilon_HH_1 t\log\big[\frac{1}{H_1\ell}\big]\Big]}{1- H_2\ell \cosh\Big[\epsilon_H H_1t \log\Big(\frac{1}{H_2\ell}\Big)\Big]}.\label{eq:Sdt}}
\item $\dot{S}_{H2}$ : This term comes from the time evolution of $H_1$ in $(c/6)\log[X(t)]$.
\dis{\dot{S}_{H2} = \frac{c}{6}\epsilon_H H_1\big( H_1\ell \big)\frac{\cosh\Big[\epsilon_HH_1 t\log\big[\frac{1}{H_1\ell}\big]\Big]}{1- H_2\ell \cosh\Big[\epsilon_H H_1t \log\Big(\frac{1}{H_2\ell}\Big)\Big]}.\label{eq:SdH2}}
\end{itemize}
 We note that whereas the negative $\dot{S}_t$   leads to the decrease of $S_{\rm mat}(\Sigma)$ in $t$ through the  explicit $t$ dependence as expected from the behavior of $S_{\rm mat}(\Sigma)$ at $t \simeq (\epsilon_H H_1)^{-1}$,  the decrease of $H_1$ in $t$ giving the positive $\dot{S}_{H1}$ and $\dot{S}_{H2}$  leads to the increase of  $S_{\rm mat}(\Sigma)$  in $t$.
  For $t$ close to $(\epsilon_H H_1)^{-1}$, $\dot{S}_t$ can easily dominate over $\dot{S}_{H1}$  as $X(t)$ in denominator of $\dot{S}_t$ is close to zero, and over $\dot{S}_{H2}$ as sinh and cosh are comparable in size but $\dot{S}_t$ contains the logarithmic enhancement factor $\log[1/(H_1\ell)]$.
   The enhancement of $\dot{S}_t$ through $1/X(t)$ does not take place when $t \ll (\epsilon_H H_1)^{-1}$ as $X(t)\lesssim 1$.
 \footnote{This is also helpful for controlling the expansion of $S_{\rm mat}(\Sigma)$ with respect to $\epsilon_H$ since the subleading terms in the expansion contain positive powers of  $1/X(t)$ which is enhanced for $X(t)$ close to $0$ (more precisely, $\epsilon_{\rm UV}^2 H_1^2$).}
  In this case, since $H_2\ell  \ll 1$, $S_{\rm mat}(\Sigma)$ is approximated as
 \dis{S_{\rm mat}(\Sigma) \simeq \frac{c}{6}\log\Big[\frac{2}{\epsilon_{\rm UV}^2 H_1H_2}\Big]\simeq \frac{c}{6}\log\Big[\frac{2}{\epsilon_{\rm UV}^2 H_1^2}\Big]-\frac{c}{6}\epsilon_H \log\Big[\frac{1}{H_1\ell}\Big],\label{eq:Smatapp}}  
 and $\dot{S}_{H1}$ can be larger than $\dot{S}_t$ and $\dot{S}_{H2}$.
 If $\dot{S}_c$ is negligibly small, $dS_{\rm mat}(\Sigma)/dt \simeq \dot{S}_{H1}$ is larger than $dS_{\rm dS}/dt$ provided 
 \dis{c> \frac{3\pi}{2}\frac{1}{G^{(3)} H_1},\label{eq:cinequal}}
or equivalently, $S_{\rm mat}(\Sigma) \gtrsim (S_{\rm dS}/2)\log[2/(\epsilon_{\rm UV}H_1)^2]$.
Then for the weak gravity case $\epsilon_{\rm UV}H_1 \simeq G^{(3)} H_1 \ll \sqrt2/e \simeq 0.5 $,  $S_{\rm mat}(\Sigma)$ is already larger than $S_{\rm dS}$ from beginning,  contradict to our setup.
 Therefore, the only way to realize the information paradox in the acceptable parameter region is allowing $c$ to depend on $t$ such that   $\dot{S}_c$ driven by $dc/dt$ is positive and dominant over ${\cal O}(\epsilon_H)$ contributions $\dot{S}_{H1}$, $\dot{S}_t$, and $\dot{S}_{H2}$.
 For instance, for the ansatz $dc/dt =\lambda\sqrt{2\epsilon_H}H_1 c$ motivated by the entropy argument for the dS swampland conjecture, $\dot{S}_c \simeq \lambda\sqrt{2\epsilon_H}H_1 S_{\rm mat}(\Sigma)$  becomes a dominant contribution to $dS_{\rm mat}(\Sigma)/dt$ as it is  ${\cal O}(\epsilon_H^{1/2})$.
 While $dS_{\rm dS}/dt$ in this case is dominated by  $(1/6)dc/dt$ term which originates from the trace anomaly, $S_{\rm mat}$ is  larger than $c/6$ by the logarithmic enhancement by $\log[1/(\epsilon_{\rm UV}^2H^2)]$, so we expect that ${\dot S}_c$ is larger than $dS_{\rm dS}/dt$.  
 
 Since time scale we consider must be much smaller than $(\epsilon_H H_1)^{-1}$, it is a good approximation to take $H_1$ to be almost constant until $S_{\rm mat}(\Sigma)$ becomes close to $S_{\rm dS}$.
  Then we can set $c\simeq c_0 {\rm exp}[\gamma \sqrt{2\epsilon_H} H_1 \Delta t]$, from which one finds that given the initial condition $S_{\rm mat}(\Sigma) \ll S_{\rm dS}$, $S_{\rm mat}(\Sigma)$ given by \eqref{eq:Smatapp}  is comparable to $S_{\rm dS}$ when 
 \dis{\Delta t =\frac{1}{\gamma}\frac{1}{\sqrt{2\epsilon_H} H_1}\Big[\log(S_{\rm dS})-\log\Big[\frac{c_0}{6}\log\Big(\frac{2}{\epsilon_{\rm UV}^2 H_1^2}\Big)\Big]\Big].\label{eq:timescale}}
 We note here that whereas $dS_{\rm dS}/dt$ is dominated by $(1/6)dc/dt$ term, $c/6$ in $S_{\rm dS}$ is required to be much smaller than the area term as $c$ is restricted to be \eqref{eq:cinequal}.
  For $\epsilon_{\rm UV} \simeq G^{(3)}$, the second term is comparable to $\log^2 S_{\rm dS}$ in size, which is much smaller than $S_{\rm dS}$ for  $S_{\rm dS} \gg 1$. 
 Then we arrive at $\Delta t\simeq (\gamma \sqrt{2\epsilon_H} H_1 )^{-1}\log(S_{\rm dS})$, which is acceptable for $\Delta t \ll (\epsilon_H H_1)^{-1}$, or $\epsilon_H^{1/2} < 1/\log(S_{\rm dS})$.

  We have seen that for the information paradox to arise in quasi-dS space, the radiation entropy which was initially much smaller than the dS entropy has to increase in time as the central charge increases, eventually exceeding the dS entropy after $\Delta t$ given by \eqref{eq:timescale}.
  It is remarkable that the situation is equivalent to that considered in the entropy argument for the dS swampland conjecture \cite{Ooguri:2018wrx} when we identify the central charge with the number of degrees of freedom. 
   That is, the entropy argument claims that when the number of low energy degrees of freedom increases exponentially  as expected from the distance conjecture, the entropy of  matter within the horizon will exceed the bound given by the dS entropy.
   Then Bousso's covariant entropy bound is violated and the backreaction of matter deforms the horizon, from which we can say  that dS space is unstable.
   In this regard, it is not strange that the time scale \eqref{eq:timescale} for the information paradox  is consistent with the time scale $(\sqrt{\epsilon_H}H)^{-1}\log(S_{\rm dS})$ at which the entropy bound is saturated by the matter entropy  \cite{Seo:2019wsh, Cai:2019dzj}.  
     If quasi-dS space has the island, the true radiation entropy $S_{\rm gen}(I \cup \Sigma)$ (which will be defined in \eqref{eq:SIR}) cannot exceed $S_{\rm dS}$ by the entanglement between the radiation and the island hence the information paradox does not arise.
   As pointed out in \cite{Teresi:2021qff}, this in turn means that in the presence of the island in quasi-dS space, the situation considered in the entropy argument for the dS swampland conjecture is not realized, which tells us that quasi-dS space is quite stable. 
  This motivates the search for the island in the inflationary quasi-dS space, which will be addressed in Sec. \ref{Sec:island}.
    
\section{Island in inflationary (quasi-)dS space}
\label{Sec:island}

  In the previous section, we argue that for the central dogma to hold without strong deformation of the quasi-dS background, we must be able to find the island within quasi-dS space.
  Meanwhile, as pointed out in \cite{Hartman:2020khs}, the island must satisfy three conditions, which forbid the island  in pure dS space.
 Small corrections by  slow-roll parameters do not change the conclusion significantly, and as we will see,  the leading correction coming from the increase of the central charge in time   makes the existence of the island more difficult.

 \subsection{Conditions on island}
 \label{Sec:islandcond}
 
 To begin with, we briefly review the conditions for the existence of the island  considered in  \cite{Hartman:2020khs}.
 Let us denote the region collecting radiation by $\Sigma$ and the island by $I$, respectively.
We can always find the region $\Sigma'$ which contains $\Sigma$ but does not intersect with $I$ ($\Sigma' \supset \Sigma$ and $\Sigma' \cap I = \emptyset$).
 Then basic inequalities on entropy like the strong subadditivity,
 \dis{S(A)+S(A') \geq S(A \cup A')+S(A \cap A'),\label{eq:SSA}}
  restrict  the region  in which the island can exist.
  To see this, we define following quantities:
  \begin{itemize}
  \item Generalized entropy of $I \cup \Sigma$
  \dis{S_{\rm gen}(I \cup \Sigma)=\frac{{\rm Area}(\partial I)}{4G}+S_{\rm mat}(I \cup \Sigma)\label{eq:SIR}}
  \item Generalized entropy of $I$
  \dis{S_{\rm gen}(I)=\frac{{\rm Area}(\partial I)}{4G}+S_{\rm mat}(I)\label{eq:SI}}
  \item Mutual information between $I$ and $\Sigma$
  \footnote{For the role of the mutual information in the island rule, see., e.g., \cite{Saha:2021ohr}}
  \dis{I(I, \Sigma)=S_{\rm gen}(I)+S_{\rm mat}(\Sigma)-S_{\rm gen}(I \cup \Sigma)=
  S_{\rm mat}(I)+S_{\rm mat}(\Sigma)-S_{\rm mat}(I \cup \Sigma)\label{eq:IIR}}
  \end{itemize}
 We also note that for the island to resolve the information paradox,
 \dis{S_{\rm gen}(I \cup \Sigma) < S_{\rm mat}(\Sigma) \label{eq:island}}
 must be satisfied.
 Among possible  $I$ satisfying \eqref{eq:island}, the island is chosen to extremize $S_{\rm gen}(I \cup \Sigma)$.

 \subsubsection{Condition 1 : violation of the area bound on $I$}
 Condition 1 is a result of two facts. 
 First, application of the island condition \eqref{eq:island} to the first equality of \eqref{eq:IIR} gives
 \dis{I (I, \Sigma) > S_{\rm gen}(I).}
 In addition, the strong subadditivity \eqref{eq:SSA} for  $A = I \cup \Sigma$ and $A'=\Sigma'$ (hence $A\cup A'=I \cup \Sigma'$ and $A \cap A'=\Sigma$) gives
 \dis{I (I, \Sigma') \geq I(I, \Sigma).}
 Choosing $\Sigma'$ to satisfy $(I\cup \Sigma')^c =\emptyset$ (hence $S(I\cup \Sigma')=S((I\cup \Sigma')^c)=0$), we obtain the relation $S_{\rm mat}({\Sigma'})=S_{\rm mat}({\Sigma'}^c)=S_{\rm mat}(I)$, resulting in  $I(I,\Sigma')=2S_{\rm mat}(I)$.
 Then above two inequalities give
 \dis{I(I,\Sigma')=2S_{\rm mat}(I) \geq I(I, \Sigma) > S_{\rm gen}(I)=S_{\rm mat}(I) +\frac{{\rm Area}(\partial I)}{4G},}
from which   condition 1
 \dis{S_{\rm mat}(I) > \frac{{\rm Area}(\partial I)}{4G} \label{eq:Cond1}}
 is obtained.
 
 \subsubsection{Condition 2 : quantum normality of $I$}
 
 Consider the deformation of $\partial I$ in the lightlike direction,  $X^\mu \to X^\mu +\lambda k^\mu$ with $k^2=0$. 
 The extremality condition $dS_{\rm gen}(I \cup \Sigma)/d\lambda=0$ in this case reads 
 \dis{\frac{d}{d\lambda}S_{\rm gen}(I)=\frac{d}{d\lambda} I(I, \Sigma),\label{eq:SIlambda}}
 where  $dS_{\rm mat}(\Sigma)/d\lambda=0$ (since $\Sigma$ is not affected by the deformation of $\partial I$) is used.
 Now suppose $\lambda \equiv \lambda_+$ is chosen such that the deformed $I$ {\it contains} $I$ before deformation : $I(\lambda_+) \subset I(\lambda_++d\lambda)$.
 From the strong subadditivity \eqref{eq:SSA} with $A=I(\lambda)\cup \Sigma$ and $A'=I(\lambda+d\lambda)$ (hence $A\cup A'=I(\lambda+d\lambda)\cup \Sigma$ and $A\cap A'=I(\lambda)$) we obtain $d I(I, \Sigma)/d\lambda_+ \geq 0$, which is equivalent to $dS_{\rm gen}(I)/d\lambda_+ \geq 0$ by \eqref{eq:SIlambda}.
 We can do the same calculation by choosing $\lambda \equiv \lambda_-$ such that the deformed $I$ {\it is contained in} $I$ before deformation, $I(\lambda_-+d\lambda) \subset I(\lambda_-)$, which results in $dS_{\rm gen}(I)/d\lambda_- \leq 0$.
 In summary, the quantum normality condition of $I$ reads 
 \dis{\pm \frac{d}{d\lambda_\pm}S_{\rm gen}(I) \geq 0.}

 \subsubsection{Condition 3 : quantum normality of $G=(I \cup \Sigma')^c$}
 
 From the strong subadditivity with $A=I(\lambda_++d\lambda)\cup \Sigma$ and $A'=I(\lambda_+)\cup \Sigma'$ (hence $A \cup A'=I(\lambda_++d\lambda)\cup \Sigma' $ and $A \cap A'=I(\lambda_+)\cup \Sigma $) we obtain 
 \dis{\frac{d}{d\lambda_+}S_{\rm mat}(I \cup \Sigma') \leq \frac{d}{d\lambda_+}S_{\rm mat}(I\cup \Sigma).}
 Using \eqref{eq:SIlambda}, one finds
 \dis{\frac{d}{d\lambda_+}S_{\rm gen}(I)&=\frac{d}{d\lambda_+}I(I, \Sigma) = \frac{d}{d\lambda_+}\big(S_{\rm mat}(I)-S_{\rm mat}(I\cup \Sigma)\big)
 \\
 &\leq \frac{d}{d\lambda_+}\big(S_{\rm mat}(I)-S_{\rm mat}(I\cup \Sigma')\big)=\frac{d}{d\lambda_+}\big(S_{\rm mat}(I)-S_{\rm mat}(G)\big),}
 or equivalently,
 \dis{0 \geq &\frac{d}{d\lambda_+}\big(S_{\rm gen}(I)-S_{\rm mat}(I)+S_{\rm mat}(G)\big) =\frac{d}{d\lambda_+}\Big(\frac{{\rm Area}(\partial I)}{4G}+S_{\rm mat}(G)\Big)
 \\
 &  =\frac{d}{d\lambda_+} S_{\rm gen}(G),}
 where for the last equality  we use the fact that $I$ and $G$ share the  boundary.
 Together with the result for $\lambda_-$ under the same step, the quantum normality condition of $G$ is written as
 \dis{\pm \frac{d}{d\lambda_\pm}S_{\rm gen}(G) \leq 0.}
 We note that condition 3 must be satisfied by any possible $G$.
 Thus, even if we find $G$ satisfying condition 3, it does not tell us the island is allowed if we can also find another $G$ violating condition 3.
 
 \subsection{Island in inflationary quasi-dS space}
 \label{Sec:islanddS}
 
 We now show that while the increase of $c$ in time gives rise to the information paradox in the inflationary quasi-dS space, it makes more difficult for the background to have the island satisfying three conditions listed in Sec. \ref{Sec:islandcond}.
 For this purpose, we keep terms up to ${\cal O}(\epsilon_H^{1/2})$ only,   by treating $H$ as a constant but taking $ (dc/dt)/c =\gamma \sqrt{2\epsilon_H}H$ into account in the analysis when we consider region I, the static patch.
 In region II, since $r$ instead of $t$ is timelike and we will consider the island on which $r$ is constant, it is convenient to consider $dc/dr$.
 Comparing with $c(\tau)$ as a function of the flat time coordinate $\tau$,
  we have $dc/dr=[(Hr)/(H^2r^2-1)] (dc/d\tau)$
 \footnote{This comes from
\dis{\frac{d}{dr}=\frac{\partial \tau}{\partial r}\frac{\partial}{\partial \tau}+\frac{\partial \rho}{\partial r}\frac{\partial}{\partial \rho}=\frac{Hr}{H^2r^2-1}\frac{\partial}{\partial \tau}- \frac{e^{-H \tau}}{H^2r^2-1}\frac{\partial}{\partial \rho}.} 
 }    
   so when we consider region II, we replace $dc/dt$ by $dc/dr=[(Hr)/(H^2r^2-1)] (\gamma \sqrt{2\epsilon_H}H c)$.
 Without the loss of generality, we consider two possible types of island :
 \begin{itemize}
 \item Type 1: The island entirely belongs to the region beyond the horizon. 
 \item Type 2: While the most part of the island belongs to the region beyond the horizon, the island extends to region I as well.
 That is, one endpoint of the island is located in region I. 
\end{itemize}  
The type 2 island   appears because the part of region I just inside the horizon can belong to the quantum system in the central dogma.
Whereas the island in this case crosses the horizon, the coordinate singularity in the static coordinates,  it does not matter as the radiation entropy depends only on the location of endpoints and the coordinate singularity is not a real singularity.
 
 Before moving onto details, we note that  the null directions parametrized by $\lambda_\pm$ are well described by the Kruskal-Szekeres null coordinates.
 From the tortoise coordinate defined by $dr_*=dr/(1-H^2r^2)$, or
   \begin{equation}
\label{eq:horwave}
r_* =
\left\{
\begin{array}{ll}
\frac{1}{2H}\log\Big[\frac{1+Hr}{1-Hr}\Big] & \text{Region~I} 
\vspace{0.5em}
\\
\frac{1}{2H}\log\Big[\frac{Hr+1}{Hr-1}\Big] & \text{Region~II} 
\vspace{0.5em}
\end{array}
\right.
\, 
,
\end{equation}
the Kruskal-Szekeres null coordinates are defined as
\dis{&U=\frac{1}{H}e^{H(t-r_*)}=\frac{1}{H}e^{Ht}\sqrt{\frac{1-Hr}{1+Hr}},
\\
&V=-\frac{1}{H}e^{-H(t+r_*)}=-\frac{1}{H}e^{-Ht}\sqrt{\frac{1-Hr}{1+Hr}},}
in region I, and
\dis{&U=\frac{1}{H}e^{H(t-r_*)}=\frac{1}{H}e^{Ht}\sqrt{\frac{Hr-1}{Hr+1}},
\\
&V=\frac{1}{H}e^{-H(t+r_*)}=\frac{1}{H}e^{-Ht}\sqrt{\frac{Hr-1}{Hr+1}},}
in region II, respectively.
They give
\dis{&\partial_U=\frac{e^{-Ht}}{2}\sqrt{\frac{1+Hr}{1-Hr}}(\partial_t-(1-H^2r^2)\partial_r),
\\
&\partial_V=\frac{e^{Ht}}{2}\sqrt{\frac{1+Hr}{1-Hr}}(\partial_t+(1-H^2r^2)\partial_r),\label{eq:dUVI}}
in region I and
\dis{&\partial_U=\frac{e^{-Ht}}{2}\sqrt{\frac{Hr+1}{Hr-1}}(\partial_t+(H^2r^2-1)\partial_r),
\\
&\partial_V=-\frac{e^{Ht}}{2}\sqrt{\frac{Hr+1}{Hr-1}}(\partial_t-(H^2r^2-1)\partial_r),\label{eq:dUVII}}
in region II.

 \subsubsection{Type 1 island}

 \begin{figure}[!t]
  \begin{center}
   \includegraphics[width=0.5\textwidth]{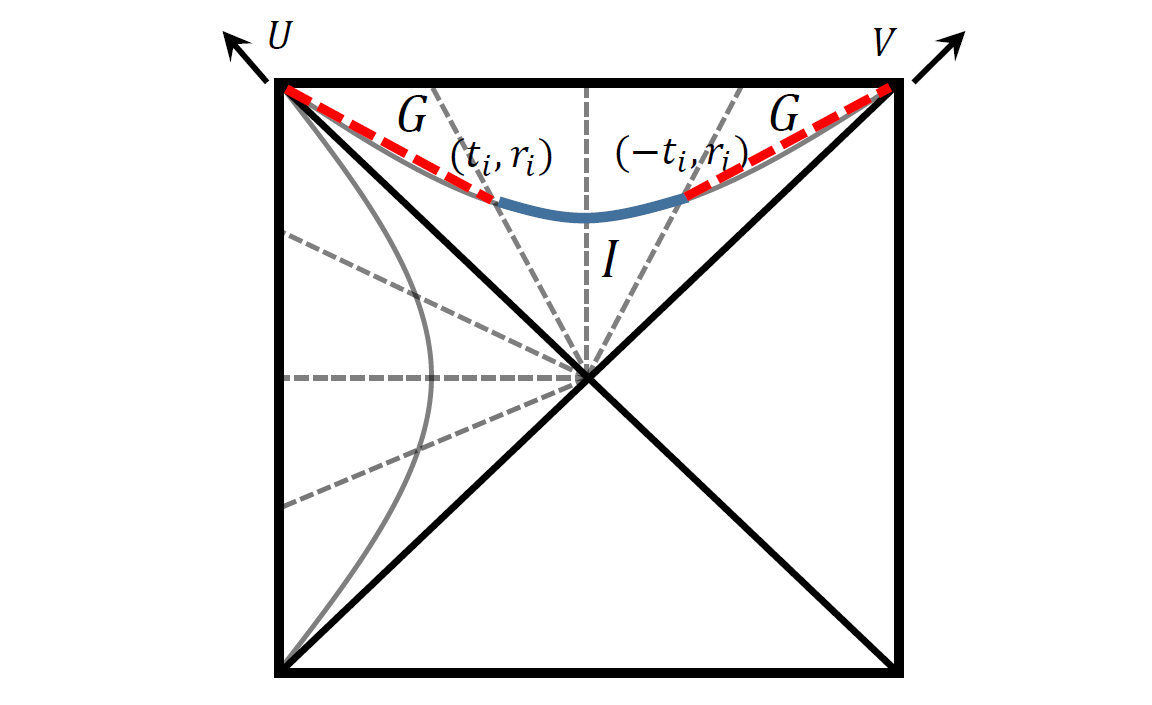}
  \end{center}
 \caption{An example of  type 1 island $I$ (thick blue line) and   region $G$ (think red dashed line). 
 The directions of the Kruskal-Szekeres coordinates $U$ and $V$ are also shown.
  }
\label{fig:island1}
\end{figure}

We consider the island on the surface $r_i=$(constant) and $t \in [-t_i, t_i]$ ($t_i>0$) in region II as depicted in Fig. \ref{fig:island1}.
 The  radiation entropy and the geometric entropy  of the island in this case are given by
 \dis{S_{\rm mat}(I)&=\frac{c}{6}\log\Big[\frac{2}{\epsilon_{\rm UV}^2H^2}\Big(1-H^2 r_i^2+(H^2r_i^2-1)\cosh[2 Ht_i]\Big)\Big],
 \\
 S_{\rm geo}(I)&=2\times \frac{\pi r_i}{2G^{(3)} }+\frac{c}{3},}
respectively, the sum of which form the generalized entropy,
\dis{S_{\rm gen}(I)=S_{\rm mat}(I)+ S_{\rm geo}(I).}

Then three conditions can be written as following inequalities:
\begin{itemize}
\item {\bf Condition 1 :}  This is simply given by
\dis{S_{\rm mat}(I) > S_{\rm geo}(I).}
\item {\bf Condition 2 :} For positive $t_i$, type 1 island $I$ expands(shrinks) as $\partial I$ moves in the direction of $U$($V$), indicating  $d/d\lambda_+=\partial/\partial U$ and $d/d\lambda_-=\partial/\partial V$.
Using \eqref{eq:dUVII}, the quantum normality condition of $I$ is written as
\dis{&\partial_U S_{\rm gen}(I) \geq 0 \Longrightarrow (\partial_{t_i} +(H^2r_i^2-1)\partial_{r_i})S_{\rm gen}(I) \geq 0
\\
& \Longrightarrow (H^2 r_i^2-1)\Big[\frac{3\pi}{c G^{(3)}H}+\frac{1}{cH}\frac{dc}{dr_i}+\frac{3 dc/dr_i}{c^2 H}S_{\rm mat}(I)\Big]+ \coth[Ht_i]+ H r_i \geq 0,
}
\dis{&\partial_V S_{\rm gen}(I)\leq 0 \Longrightarrow -(\partial_{t_i} -(H^2r_i^2-1)\partial_{r_i})S_{\rm gen}(I) \leq 0
\\
&\Longrightarrow -(H^2r_i^2-1)\Big[\frac{3\pi}{c G^{(3)}H}+\frac{1}{cH}\frac{dc}{dr_i}+\frac{3 dc/dr_i}{c^2 H}S_{\rm mat}(I)\Big]+\coth[Ht_i]- H r_i \geq 0. \label{eq:pVSn}}
\item {\bf Condition 3 :}  When we take  $\Sigma'$ to be region I, the region $G$ can be chosen to be the subregion of the $r_i=$(constant) surface in region II complement to $I$ (the thick red dashed line in Fig. \ref{fig:island1}).
Then $G$ consists of two  disconnected regions, each of which shares the  boundary with $I$.
The generalized entropy of $G$ is given by
\dis{S_{\rm gen}(G)&=\frac{\pi r_i}{G^{(3)}}+\frac{c}{3}+S_{\rm mat}(G)
\\
&=\frac{\pi r_i}{G^{(3)}}+\frac{c}{3}+\frac{c}{3}\log\Big[\frac{2}{\epsilon_{\rm UV}^2H^2}\Big(1-H^2 r_i^2+(H^2 r_i^2-1)\cosh[H(T_0-t_i)]\Big)\Big],}
where $T_0$ is taken to be infinity, from which the quantum normality condition of $G$ is written as
\dis{&\partial_U S_{\rm gen}(G) \leq 0
\\
&\Longrightarrow  (H^2r_i^2-1)\Big[\frac{3\pi}{c G^{(3)}H}+\frac{1}{cH}\frac{dc}{dr_i}+\frac{3 dc/dr_i}{c^2 H}S_{\rm mat}(G)\Big]-  \coth[\frac{H}{2}(T_0-t_i)]+ 2 H r_i \leq 0,
\label{eq:pUSGn}}
\dis{&\partial_V S_{\rm gen}(G)\geq 0
\\
&\Longrightarrow  -(H^2r_i^2-1)\Big[\frac{3\pi}{c G^{(3)}H}+\frac{1}{cH}\frac{dc}{dr_i}+\frac{3 dc/dr_i}{c^2H}S_{\rm mat}(G)\Big]-\coth[\frac{H}{2}(T_0-t_i)]- 2H r_i \leq 0.}
\end{itemize}

 \begin{figure}[!t]
  \begin{center}
   \includegraphics[width=0.4\textwidth]{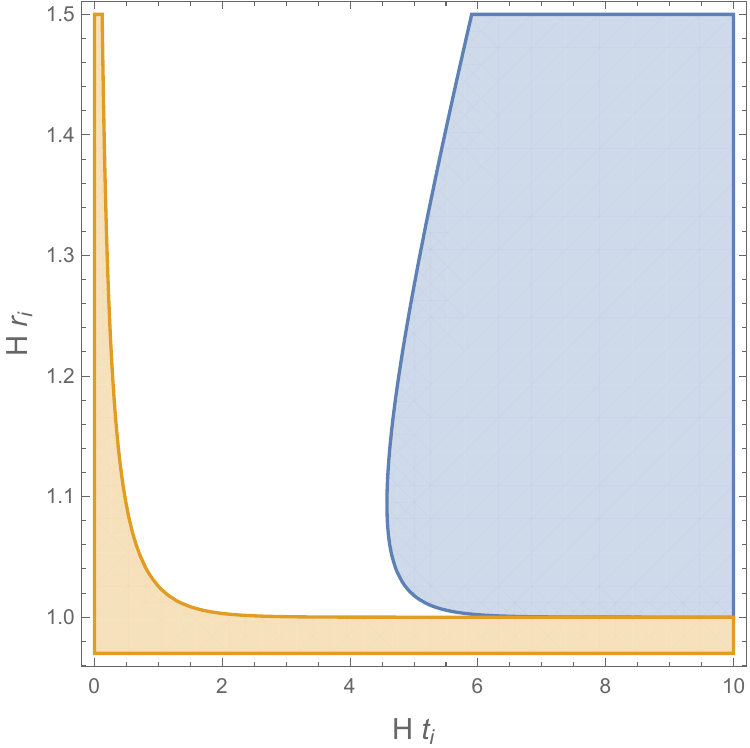}
   \includegraphics[width=0.4\textwidth]{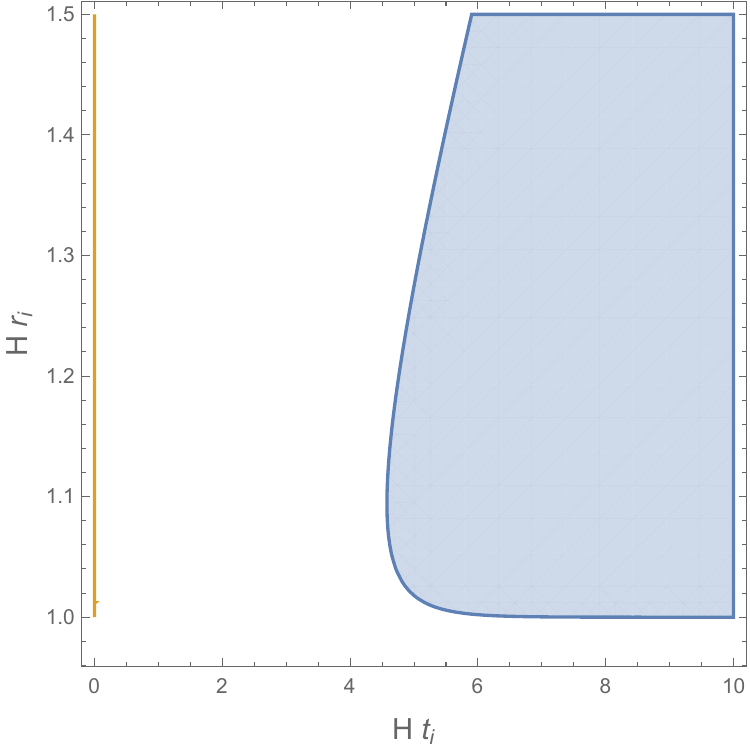}
  \end{center}
 \caption{Regions satisfying three conditions. We set $cG^{(3)}H/(6\pi) =0.09$ and $G^{(3)}H=\epsilon_{\rm UV}H=0.1$.
 (Left) : Constant central charge ($dc/dr_i=0$) case. Regions allowed by condition 1 and $\partial_V S_{\rm gen}(I) \leq 0$ in condition 2  are colored in blue and orange, respectively.
  They do not overlap. 
  While $\partial_U S_{\rm gen}(G) \leq 0$ in condition 3 is not satisfied in any region, other conditions are satisfied in the whole region $Hr_i >1$.
 (Right) : Evolving central charge case with  $(dc/dr_i)/(c H)=0.5[(Hr_i)/(H^2r_i^2-1)]$. While condition 1 is unchanged and $\partial_U S_{\rm gen}(I) \geq 0$ in condition 2   still holds in whole region  $Hr_i >1$, the region satisfying $\partial_V S_{\rm gen}(I) \leq 0$ in condition 2 shrinks.
 Moreover, $\partial_U S_{\rm gen}(G) \leq 0$ in condition 3 is not satisfied in any region. 
  }
\label{fig:island1plot}
\end{figure}

We first consider the ${\cal O}(\epsilon_H^0)$ effects only by setting the  ${\cal O}(\epsilon_H^{1/2})$ correction $(dc/dr_i)/(c H)$ zero.
Condition 1 is easily satisfied for $H t_i \gg 1$.
The region $Ht_i\lesssim {\cal O}(1)$ is allowed by condition 1 when $c G^{(3)}H/(6\pi)>1$ but  this  is not an appropriate parameter value for $dc/dr_i=0$ as $S_{\rm mat}(\Sigma)$ in this case can be larger than $S_{\rm dS}$, i.e., $(c/6)\log[2/(\epsilon_{\rm UV}^2 H^2)] > \pi/(2 G^{(3)}H)$.
We allow the case of $c G^{(3)}H/(6\pi)>1$  as a consequence of the increase of $c$ by nonzero $dc/dr_i$ only, in which  the information paradox arises.
For condition 2,  $\partial_U S_{\rm gen} (I) \geq 0$, or equivalently,
\dis{Hr_i \geq-\Big(\frac{c G^{(3)} H}{6\pi}\Big)+\sqrt{1-2\Big(\frac{c G^{(3)} H}{6\pi}\Big)\coth (Ht_i)+\Big(\frac{c G^{(3)} H}{6\pi}\Big)^2}\label{eq:ineq1}}
is approximated as
\dis{Hr_i \gtrsim -\Big(\frac{c G^{(3)} H}{6\pi}\Big) +\Big| 1-\Big(\frac{c G^{(3)} H}{6\pi}\Big) \Big|}
for $H t_i \gg 1$ ($\coth (Ht_i)\simeq 1$), which is trivially satisfied in the region $Hr_i>1$ regardless of the size of the `strength of gravitation' $cG^{(3)}H/(6\pi)$.
On the other hand,  $\partial_V S_{\rm gen} (I) \leq 0$ is written as
\dis{Hr_i \leq-\Big(\frac{c G^{(3)} H}{6\pi}\Big)+\sqrt{1+2\Big(\frac{c G^{(3)} H}{6\pi}\Big)\coth (Ht_i)+\Big(\frac{c G^{(3)} H}{6\pi}\Big)^2},\label{eq:ineq2}}
which is approximated as 
\dis{Hr \lesssim -\Big(\frac{c G^{(3)} H}{6\pi}\Big) +\Big| 1+\Big(\frac{c G^{(3)} H}{6\pi}\Big) \Big| = 1}
for $Ht_i \gg 1$, excluding the region $Hr_i >1$ we are considering.
Finally, from $\coth[(H/2)(T_0-t_i)] \to 1$ as $T_0\to \infty$,  condition 3 reads
\dis{&Hr_i \leq -\frac{c G^{(3)}H}{3\pi}+\sqrt{1+\frac{c G^{(3)}H}{3\pi}+\Big(\frac{c G^{(3)}H}{3\pi}\Big)^2},
\\
&Hr_i\geq -\frac{c G^{(3)}H}{3\pi}+\sqrt{1-\frac{c G^{(3)}H}{3\pi}+\Big(\frac{c G^{(3)}H}{3\pi}\Big)^2}.\label{eq:ineq3}}
Since RHS of both inequalities are smaller than $1$, the first one excludes while the second one allows the whole region  $Hr_i >1$.
Our analysis so far shows that the region satisfying three conditions simultaneously does not exist when $c$ is constant thus type 1 island  is not allowed in perfect dS space.
 In the left panel of Fig. \ref{fig:island1plot} we show our conclusion explicitly for specific choice of parameters.

Taking nonzero $(dc/dr_i)/(c H) \sim {\cal O}(\epsilon_H^{1/2})$ into account makes the situation even   worse, as can be found in the right panel of Fig. \ref{fig:island1plot}.
\footnote{We note that while $S_{\rm geo}$ contains a term $c/6$ reflecting the trace anomaly, it does not play the crucial role as it is suppressed compared to $S_{\rm mat}$ as the latter has a logarithmic enhancement.}
 For  $Hr_i > 1$, both $S_{\rm mat}(I)$ and $S_{\rm mat}(G)$ are positive unless $Hr_i$ is very close to $1$.
 In this case,  the addition of positive term $(H^2r_i^2-1)[(3cH)^{-1}dc/dr_i+(dc/dr_i)c^{-1}S_{\rm mat}(I)]$ to $\partial_{U, V}S_{\rm gen}(I)$ in condition 2 and $(H^2r_i^2-1)[(3cH)^{-1}dc/dr_i+(dc/dr_i)c^{-1}S_{\rm mat}(G)]$ to $\partial_{U, V}S_{\rm gen}(G)$ in condition 3  make  the inequalities $\partial_{U}S_{\rm gen}(I) \geq 0$ and $\partial_{V}S_{\rm gen}(G) \geq 0$ easier but the inequalities $\partial_{V}S_{\rm gen}(I) \leq 0$ and $\partial_{U}S_{\rm gen}(G) \leq 0$ more difficult to be satisfied. 
 This also can  be seen by noticing that  RHS of \eqref{eq:ineq2} coming from $\partial_{V}S_{\rm gen}(I) \leq 0$ in condition 2 is larger than $1$ (since $\coth(Ht_i)>1$), and decreases as $(3\pi)/(c G^{(3)}H)$ becomes larger.
   In the presence of positive $dc/dr_i$, $(3\pi)/(c G^{(3)}H)$ in condition 2 is modified as
\dis{\frac{3\pi}{c G^{(3)}H}+\frac{1}{cH}\frac{dc}{dr_i}+\frac{3 dc/dr_i}{c^2 H}S_{\rm mat}(I),}
so the addition of positive term $(cH)^{-1
}dc/dr_i+[(3dc/dr_i)/(c^2H)]S_{\rm mat}(I)$ may be regarded as the increase of $(3\pi)/(c G^{(3)} H)$ effectively.
This makes RHS of \eqref{eq:ineq2} smaller, which means that the inequality becomes more restrictive.
 On the other hand, all the RHS of other inequalities \eqref{eq:ineq1} and \eqref{eq:ineq3} get larger as $(3\pi)/(c G^{(3)}H)$ increases.
 However, observing changes of these bounds in the presence of the positive $dc/dr_i$ is not meaningful since all these values  are smaller than $1$, while we are considering $Hr_i$ larger than $1$.

 The increase of $c$ driven by nonzero $dc/dr_i$ leads to the information paradox as $S_{\rm mat}(\Sigma)$ becomes larger than $S_{\rm dS}$. 
 In this case,  condition 1 is satisfied even in  the region $Ht_i \lesssim 1$, which may allow the overlap with the region satisfying $\partial_V S_{\rm gen}(I) \leq 0$.
 Indeed, since $(cG^{(3)}H)/(6\pi)$ is no longer much less than $1$, $\partial_V S_{\rm gen}(I)\leq 0$ in condition 2 reads
 \dis{Hr_i \lesssim \coth[Ht_i]-(H^2r_i^2-1)\Big(\frac{1}{cH}\frac{dc}{dr_i}+\frac{3dc/dr_i}{c^2H}S_{\rm mat}(I)\Big).}
 While this can be satisfied in some part of the region $Hr_i>1$ as $\coth[Ht_i]$ is larger than $1$, the negative term containing $dc/dr_i$ still makes the inequality restrictive even if $(3\pi)/(cG^{(3)}H)$ is suppressed.
 Moreover, $\partial_U S_{\rm gen}(G)\leq 0$ in condition 3 becomes
 \dis{Hr_i \lesssim \frac12-(H^2r_i^2-1)\Big(\frac{1}{2cH}\frac{dc}{dr_i}+\frac{3dc/dr_i}{2c^2H}S_{\rm mat}(G)\Big),} 
 in which RHS is evidently less than $1$, excluding the whole region $Hr_i>1$.
 This shows that type 1 island does not exist in the presence of the information paradox.

 \begin{figure}[!t]
  \begin{center}
   \includegraphics[width=0.4\textwidth]{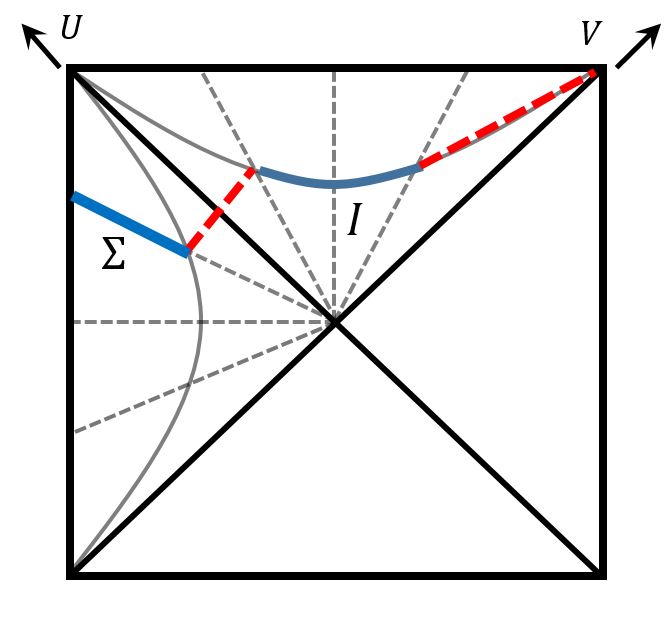}
  \end{center}
 \caption{The choice of Cauchy slice, the union of $\Sigma$, $I$, and the red dashed line.
  }
\label{fig:island15}
\end{figure}

We can also understand the nonexistence of type 1 island in a more direct way.
Assuming the quantum state on the Cauchy slice to be pure, $S_{\rm mat}(I \cup \Sigma)$ is identified with $S_{\rm mat}((I \cup \Sigma)^c)$, the matter entropy on the red dashed line in Fig. \ref{fig:island15}.
Adding the boundary area to this, we obtain
\dis{S_{\rm gen}(I \cup \Sigma) = &\frac{\pi r}{G^{(3)}}+2\frac{\pi r_i}{G^{(3)}}+c
\\
&+\frac{c}{6}\log\Big[\frac{2}{\epsilon_H^2H^2}\big(1-H^2 r r_i
\sqrt{(1-H^2r^2)(H^2r_i^2-1)}\sinh[H(t-t_1)]\big)\Big]
\\
&+\frac{c}{6}\log\Big[\frac{2}{\epsilon_H^2H^2}\big(1-H^2r_i^2+(H^2r_i^2-1)\cosh[H(T_0-t_i)]\big)\Big],}
where $r$ is the coordinate of the right end of $\Sigma$.
While the island is required to be chosen to extremize $S_{\rm gen}(I \cup \Sigma)$ in both $t_i$ and $r_i$ directions, 
\dis{\frac{\partial S_{\rm gen}(I \cup \Sigma)}{\partial t_i}=-\frac16 c H -\frac{cH}{6}\frac{\sqrt{(1-H^2r^2)(H^2r_i^2-1)}\cosh H(t-t_1)}{1-H^2 r r_i+\sqrt{(1-H^2r^2)(H^2r_i^2-1)}\sinh[H(t-t_i)]},}
where the limit $T_0 \to \infty$ is taken, is always negative, which means that $S_{\rm gen}(I \cup \Sigma)$ is not extremized in the $t_i$ direction.

\subsubsection{Type 2 island}

 \begin{figure}[!t]
  \begin{center}
   \includegraphics[width=0.5\textwidth]{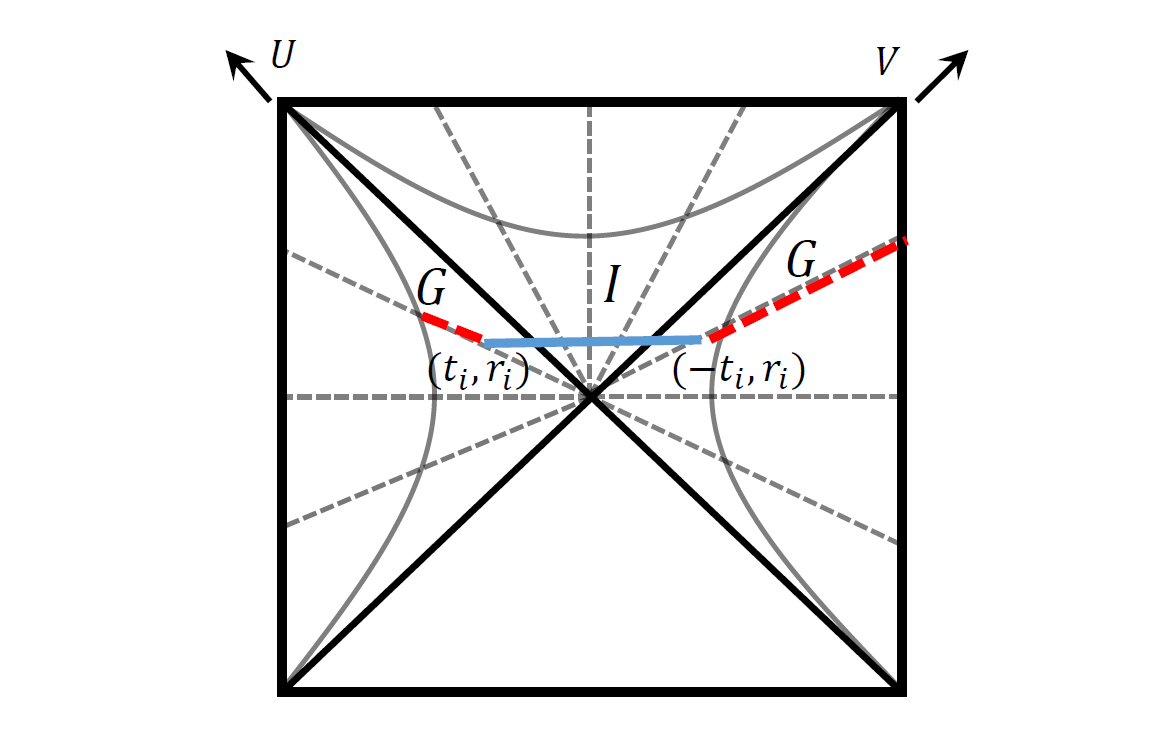}
  \end{center}
 \caption{An example of type 2 island $I$ (thick blue line) and   region $G$ (think red dashed line). 
 The left part of $G$ is so narrow that  its contribution to $S_{\rm mat}(G)$ is negligible. 
  }
\label{fig:island2}
\end{figure}

 In order to collect the radiation in (quasi-)dS space, we choose the hypersurface $\Sigma$ in the part of region I in which gravity is treated as a nondynamical background.
 The quantum gravity region extends over the rest part of region I as well as the region beyond the horizon.
 Since the island can exist in any part of quantum gravity region, we can consider  type 2 island as depicted in Fig. \ref{fig:island2} : whereas the most part of the island belongs to  region II, the island also extends to  region I and III.
 Let the static coordinates of two endpoints are given by $(t_i, r_i)$ (in region I) and $(-t_i, r_i)$ (in region III), respectively.
 Here the right end of the island in region III is the reflection of the left end of the island in region I with respect to the line $U=V$ in Fig. \ref{fig:island2},   $X^{0, 1} \to X^{0, 1}$ and $X^2 \to -X^2$. or equivalently,  $\sigma \to\sigma$ and $\theta \to \pi-\theta$ in the conformal coordinates.
 Thus,   the embeddings in  three-dimensional Minkowski space   are given by
 \dis{X_R^0=H^{-1}\sqrt{1-H^2r_i^2}\sinh(Ht_i),\quad X_R^1=r_i,\quad X_R^2=-H^{-1}\sqrt{1-H^2r_i^2}\cosh(Ht_i),}  
 for the right end of the island, and 
 \dis{X_L^0=H^{-1}\sqrt{1-H^2r_i^2}\sinh(Ht_i),\quad X_L^1=r_i,\quad X_L^2= H^{-1}\sqrt{1-H^2r_i^2}\cosh(Ht_i),}
 for the left end of the island, respectively.
 From this, the radiation entropy of the island is written as  
 \dis{S_{\rm mat}(I)&=\frac{c}{6}\log\Big[\frac{2}{\epsilon_{\rm UV}^2H^2}(1-H^2 X_L\cdot X_R)\Big]
 \\
  &=\frac{c}{6}\log\Big[\frac{2}{\epsilon_{\rm UV}^2 H^2}\Big(1-H^2r_i^2+(1-H^2r_i^2)\cosh[2 Ht_i]\Big)\Big].}
Adding this to the geometric entropy,
\dis{S_{\rm geo}(I)=2\times \frac{\pi r_i}{2G^{(3)}}+\frac{c}{3}}
we obtain the generalized entropy of the island,
\dis{S_{\rm gen}(I)=S_{\rm mat}(I)+ S_{\rm geo}(I).}

\begin{itemize}
\item {\bf Condition 1 :}  This is simply given by
\dis{S_{\rm mat} (I)> S_{\rm geo}(I).}
\item {\bf Condition 2 :}  Type 2 island $I$ expands(shrinks) as $\partial I$ moves in the direction of $U$($V$), indicating  $d/d\lambda_+=\partial/\partial U$ and $d/d\lambda_-=\partial/\partial V$.
From \eqref{eq:dUVI}, we obtain the quantum normality condition on $I$,
\dis{&\partial_U S_{\rm gen}(I) \geq 0 \Longrightarrow (\partial_{t_i} -(1-H^2r_i^2)\partial_{r_i})S_{\rm gen}(I) \geq 0,
\\
& \Longrightarrow \frac{1}{cH}\frac{dc}{dt_i}+\frac{3 dc/dt_i}{c^2 H}S_{\rm mat}(I)-(1-H^2 r_i^2)\frac{3\pi}{c G^{(3)}H}+ \tanh[Ht_i]+H r_i \geq 0
\label{eq:T2Cond2}}
\dis{&\partial_V S_{\rm gen}(I)\leq 0 \Longrightarrow (\partial_{t_i} +(1-H^2r_i^2)\partial_{r_i})S_{\rm gen}(I) \leq 0
\\
&\Longrightarrow  \frac{1}{cH}\frac{dc}{dt_i}+\frac{3 dc/dt_i}{c^2 H}S_{\rm mat}(I)+(1-H^2r_i^2)\frac{3\pi}{c G^{(3)} H}+\tanh[Ht_i]- H r_i \leq 0.\label{eq:T2Cond2v}}
\item {\bf Condition 3 :} We consider the following choice of $G$ as shown in Fig. \ref{fig:island2}.
First, in order to avoid the overlap with $\Sigma$ extending over the most part of region I, the part of $G$ in region I occupies only a narrow interval of length $\simeq \epsilon_{\rm UV}$ thus the radiation entropy in this part is negligible.
While one of boundaries of this part is not shared by $I$, it is not relevant to our discussion   since we are interested in the variation of $G$ under the deformation of boundary shared by $I$.
Meanwhile, another part of $G$ lies on $-t_i=$(constant) surface in region III such that the endpoints are given by $(-t_i, 0)$ and $(-t_i, r_i)$.
Then up to the addition of irrelevant term,  the generalized entropy of $G$ is given by
\dis{S_{\rm gen}(G)= \frac{c}{6}\log\Big[\frac{2}{\epsilon_{\rm UV}^2 H^2}\Big(1-\sqrt{1-H^2 r_i^2}\Big)\Big]+ S_{\rm geo}(I).}
Then the quantum normality condition on G is written as
\footnote{More precisely, the part of $G$ we are considering is in region III in which $I$ expands(shrinks) in the direction of $V(U)$ with $\partial_{U/V}\propto -\partial_t \pm (1-H^2r^2)\partial_r$.
However, if we set $c$ to increase toward the future infinity, $dc/dt_i<0$ so $-dc/dt_i$ in region III can be replaced by the positive $dc/dt_i$ in region I giving \eqref{eq:T2Cond3}. }
\dis{&\partial_U S_{\rm gen}(G) \leq 0
\\\
& \Longrightarrow \frac{2}{cH}\frac{dc}{dt_i}+ \frac{6 dc/dt_i}{c^2 H}S_{\rm mat}(G)-(1-H^2r_i^2)\frac{6\pi}{c G^{(3)}H}-\frac{\sqrt{1-H^2r_i^2}}{1-\sqrt{1-H^2r_i^2}} H r_i \leq 0,
\\
&\partial_V S_{\rm gen}(G)\geq 0
\\
&\Longrightarrow \frac{2}{cH}\frac{dc}{dt_i}+\frac{6 dc/dt_i}{c^2 H}S_{\rm mat}(G)+ (1-H^2r_i^2)\frac{6\pi}{c G^{(3)}H}+ \frac{\sqrt{1-H^2r_i^2}}{1-\sqrt{1-H^2r_i^2}} H r_i \geq 0.\label{eq:T2Cond3}}
We note that whereas two conditions are equivalent for $dc/dt_i=0$, when $dc/dt_i>0$ the first one gives more stringent bound.
\end{itemize}

 \begin{figure}[!t]
  \begin{center}
   \includegraphics[width=0.4\textwidth]{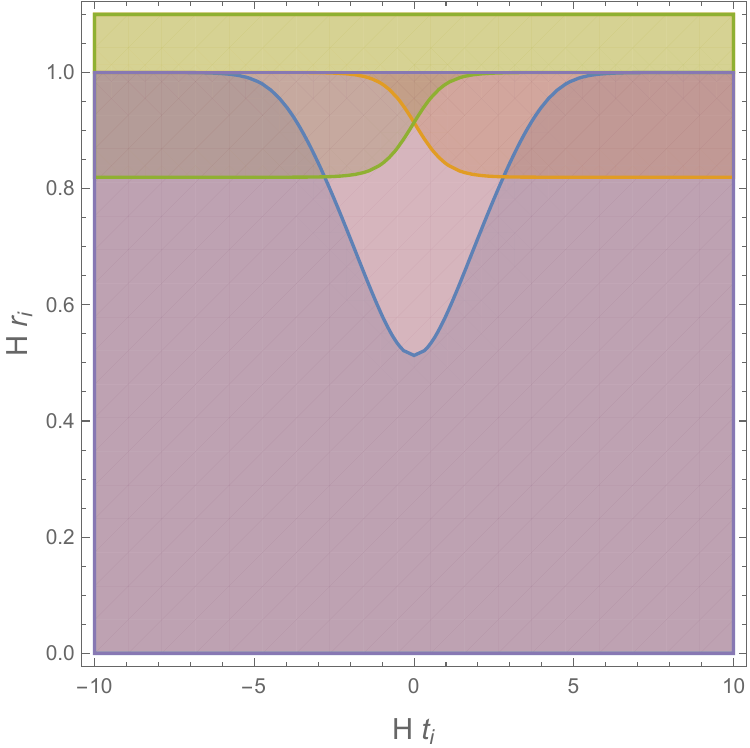}
   \includegraphics[width=0.4\textwidth]{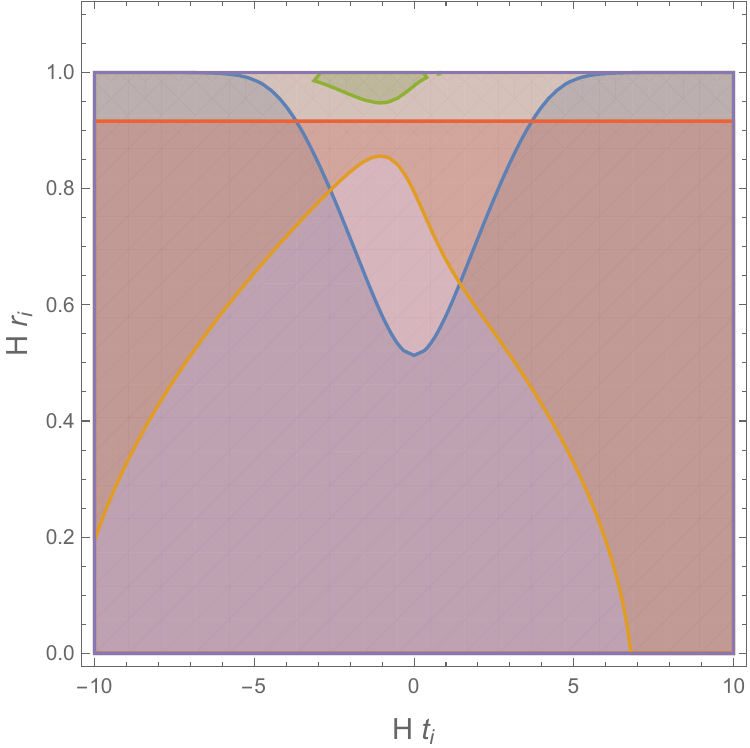}
  \end{center}
 \caption{Regions satisfying three conditions. We set $cG^{(3)}H/(6\pi) =0.09$ and $G^{(3)}H=\epsilon_{\rm UV}H=0.1$.
 (Left) : Constant central charge ($dc/dt_i=0$) case. 
 Regions satisfying condition 1, $\partial_U S_{\rm gen}(I) \geq 0$ in condition 2, $\partial_V S_{\rm gen}(I) \leq 0$ in condition 2  and condition 3 are colored in blue, orange, green, and red, respectively. 
 The region allowed by all three conditions does not exist, especially because $\partial_U S_{\rm gen}(I) \geq 0$ and $\partial_V S_{\rm gen}(I) \leq 0$ in condition 2 are satisfied simultaneously only for $Ht_i\simeq 0$ and $Hr_i \lesssim 1$, which is excluded by condition 1. 
 (Right) : Evolving central charge case with $(dc/dt_i)/(c H)=0.5$. 
 Regions satisfying condition 1, $\partial_U S_{\rm gen}(I) \geq 0$ in condition 2, $\partial_V S_{\rm gen}(I) \leq 0$ in condition 2,  $\partial_U S_{\rm gen}(G) \leq 0$ in condition 3, and $\partial_V S_{\rm gen}(G) \geq 0$ in condition 3 are colored in blue, orange, green, red, and purple,  respectively.
  The nonzero $dc/dt_i$ makes two inequalities in condition 3 different, as can be found from the reduction of the region satisfying $\partial_U S_{\rm gen}(G) \leq 0$.
 While the region satisfying  $\partial_U S_{\rm gen}(I) \geq 0$ in condition 2 is enhanced, the region satisfying $\partial_V S_{\rm gen}(I) \leq 0$ in condition 2 shrinks and becomes exclusive to $\partial_U S_{\rm gen}(G) \leq 0$.
The region satisfying all three conditions does not exist.
  }
\label{fig:island2plot}
\end{figure}

We first consider the ${\cal O}(\epsilon_H^0)$ effects, in which $dc/dt_i$ is neglected.
In addition, for the time being, we take $t_i$ to be positive.
In fact, whereas $t_i$ appears in condition 1 and 2, $S_{\rm mat}(I)$ is even in $t_i$ so condition 1 is just a condition on $|t_i|$.
For  condition 2, when $dc/dt_i$ term is neglected, the only difference between $\partial_U S_{\rm gen} (I) \geq 0$ and $\partial_V S_{\rm gen} (I) \leq 0$  is the sign of the $\tanh(Ht_i)$ term, hence $\partial_U S_{\rm gen} (I) \geq 0$ for positive $t_i$ is equivalent to $\partial_V S_{\rm gen} (I) \leq 0$ for negative $t_i$ and vice versa.
Just like the case of type 1 island,  condition 1 is easily satisfied when $Ht_i \gg 1$.
The inequality $\partial_U S_{\rm gen} (I) \geq 0$ in condition 2 reads
\dis{Hr \geq-\Big(\frac{c G^{(3)} H}{6\pi}\Big)+\sqrt{1-2\Big(\frac{c G^{(3)} H}{6\pi}\Big)\tanh (Ht_i)+\Big(\frac{c G^{(3)} H}{6\pi}\Big)^2}.\label{eq:T2cond2(1)}}
Here RHS is smaller than $1$ since $\tanh(Ht_i)<1$, and for $Ht_i\gg 1$, the inequality becomes
\dis{Hr \gtrsim -\Big(\frac{c G^{(3)} H}{6\pi}\Big) +\Big| 1-\Big(\frac{c G^{(3)} H}{6\pi}\Big) \Big|.}
For the weak gravity case,  $(c G^{(3)} H )/(6\pi) \ll 1$, this condition becomes $H r_i \gtrsim 1-(c G^{(3)} H )/(3\pi)$ which allows the region close to the horizon.
 For the strong gravity case, $(c G^{(3)} H )/(6\pi)  \gtrsim 1$, the condition becomes $H r_i \gtrsim -1$, which is trivially satisfied in all region inside the horizon.
 In addition, the region satisfying condition 1 is extended as well. 
 This will be considered later with nonzero $dc/dt_i$ effect to discuss the information paradox as  $S_{\rm mat}(\Sigma)$ in this case is no longer smaller  than $S_{\rm dS}$.
 Next, the  condition $\partial_V S_{\rm gen} (I) \leq 0$ is written as
\dis{Hr_i \geq -\Big(\frac{c G^{(3)} H}{6\pi}\Big)+\sqrt{1+2\Big(\frac{c G^{(3)} H}{6\pi}\Big)\tanh (Ht_i)+\Big(\frac{c G^{(3)} H}{6\pi}\Big)^2}.}
For $Ht_i \gg 1$, this condition reads 
\dis{Hr_i \gtrsim -\Big(\frac{c G^{(3)} H}{6\pi}\Big) +\Big| 1+\Big(\frac{c G^{(3)} H}{6\pi}\Big) \Big| \simeq 1,}
which allows the region around the horizon only.
 Since two inequalities $\partial_U S_{\rm gen} (I) \geq 0$ and  $\partial_V S_{\rm gen} (I) \leq 0$ can be interchanged by  $t_i \to -t_i$, the regions allowed by them can overlap for $t_i \simeq -t_i$, i.e., $Ht_i\simeq 0$. 
 For the  weak coupling case  $(c G^{(3)} H )/(6\pi) \ll 1$, the value of $Hr_i$ satisfying two inequalities simultaneously for $Ht_i \simeq 0$ is close to $1$, which is easily excluded by condition 1. 
 More concretely, as $\cosh(2Ht_i)\simeq 1$ and $1-H^2 r_i^2 \ll 1$, $S_{\rm mat}(I)$ becomes much smaller than $S_{\rm mat}(\Sigma)\simeq (c/6)\log[2/(\epsilon_{\rm UV}^2H^2)]$  which is required to be bounded by $S_{\rm dS}=\pi/(2G^{(3)} H)$.
 But at the same time, since $r_i$ is close to $H^{-1}$, $S_{\rm geo}(I)$ becomes close to $S_{\rm dS}$, leading to $S_{\rm mat}(I)<S_{\rm geo}(I)$, contradict to condition 1.
 Indeed, when $Ht_i\simeq 0$, the value of $Hr_i$ satisfying two inequalities in condition 2 is given by
 \dis{Hr_i \simeq -\Big(\frac{c G^{(3)} H}{6\pi}\Big)+\sqrt{1+\Big(\frac{c G^{(3)} H}{6\pi}\Big)^2}.}
 Then for the weak gravity case $(c G^{(3)}H)/(6\pi)\ll 1$, we have 
 \dis{S_{\rm mat}(I)\simeq \frac{c}{6}\log\Big[\frac{8}{\epsilon_{\rm UV}^2H^2}\frac{c G^{(3)} H}{6\pi}\Big]\simeq \frac{c}{6}\log\Big[\frac{8 c}{(6\pi)^2}\frac{6\pi}{  G^{(3)} H}\Big],}
 where we use $\epsilon_{\rm UV} \simeq G^{(3)}$.
  Comparing this with  $S_{\rm geo}(I)\simeq \pi (G^{(3)}H)^{-1}$, we find that $S_{\rm geo}(I)$ is larger than $S_{\rm mat}(I)$ since they are  linear and logarithmic in large value   $(6\pi)/(c G^{(3)}H)$, respectively.
 This is not consistent with condition 1.
  Finally, one finds that for $dc/dt_i=0$, two inequalities in condition 3 give the same condition and trivially satisfied.
 Features of three conditions discussed so far are summarized in the left panel of Fig. \ref{fig:island2plot}, showing that the region satisfying all three conditions does not exist.

When we take the nonzero, positive $dc/dt_i$ into account, as shown in the right panel of Fig. \ref{fig:island2plot},   conditions $\partial_U S_{\rm gen} (G) \leq 0$ and $\partial_V S_{\rm gen} (I) \leq 0$ become more difficult to be satisfied, just like the case of island 1.
In particular, the region allowed by $\partial_U S_{\rm gen} (G) \leq 0$ is more restricted to $Ht_i\simeq 0$ and $Hr_i\simeq 1$, in which condition 1 is violated as explained above.

 \begin{figure}[!t]
  \begin{center}
   \includegraphics[width=0.4\textwidth]{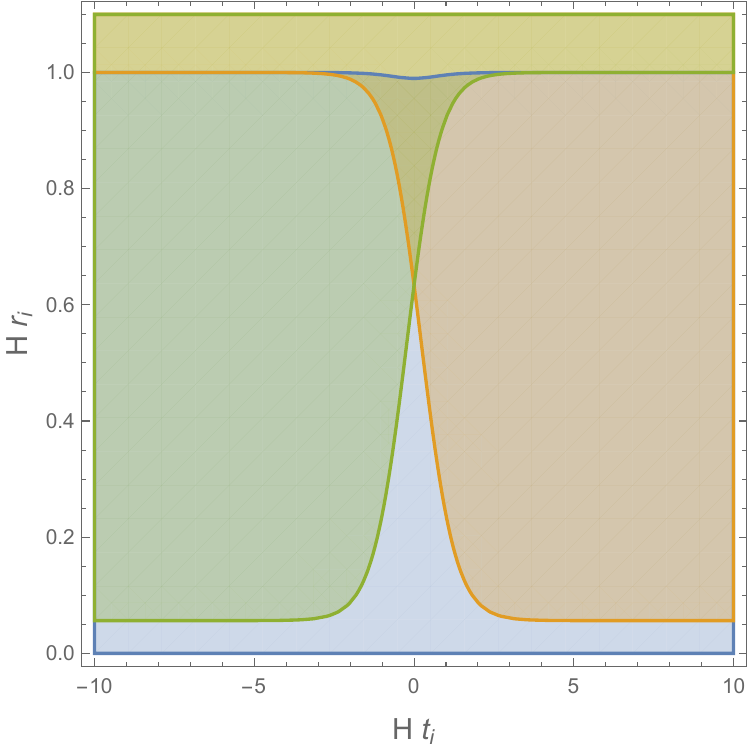}
   \includegraphics[width=0.4\textwidth]{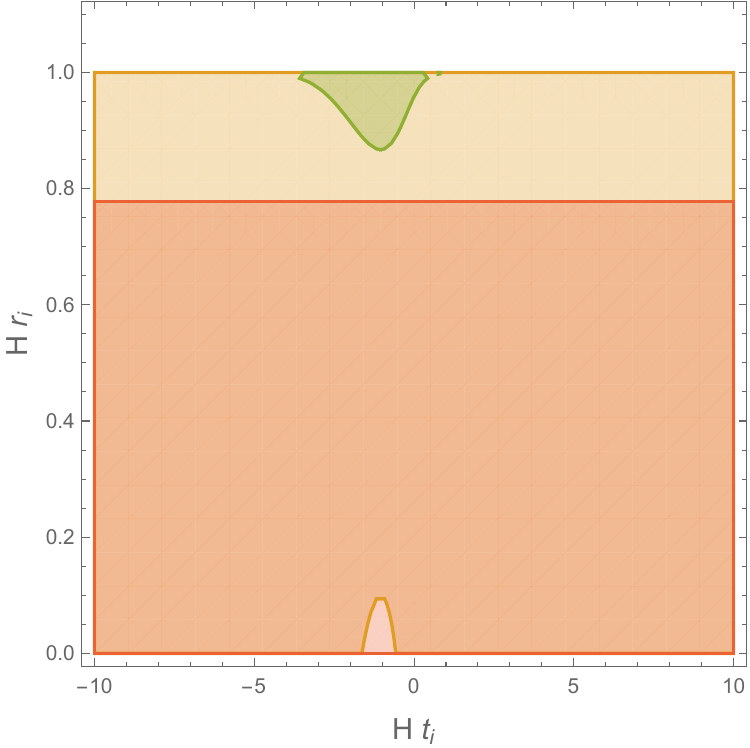}
  \end{center}
 \caption{Regions satisfying three conditions for $cG^{(3)}H/(6\pi) =5\times 0.09$ and $G^{(3)}H=\epsilon_{\rm UV}H=0.1$.
 (Left) : Constant central charge ($dc/dt_i=0$) case. 
 Regions satisfying condition 1, $\partial_U S_{\rm gen}(I) \geq 0$ in condition 2, and $\partial_V S_{\rm gen}(I) \leq 0$ in condition 2  are colored in blue, orange, and green respectively. 
The region allowed by condition 3 is not drawn as it  is trivially satisfied in the whole region $Hr_i<1$.
 The region allowed by all the conditions can be found for $H|t_i|<2$ and $Hr_i>0.6$.
 (Right) : Evolving central charge case with $(dc/dt_i)/(c H)=0.5$. 
 Regions satisfying  $\partial_U S_{\rm gen}(I) \geq 0$ in condition 2,  $\partial_V S_{\rm gen}(I) \leq 0$ in condition 2, and $\partial_U S_{\rm gen}(G) \leq 0$ condition 3  are colored in orange, green and red,   respectively. 
 Regions allowed by $\partial_V S_{\rm gen}(I) \leq 0$ and $\partial_U S_{\rm gen}(G) \leq 0$ do not overlap.
 The region satisfying condition 1 is not shown since it is the same as the case of $dc/dr_i=0$ and satisfied in the whole region $H r_i<1$ except for tiny region around $Ht_i\simeq 0$ and $Hr_i\simeq 1$.
 The region  allowed by   $\partial_V S_{\rm gen}(G) \geq 0$ in condition 3 is not shown as well since it is trivially satisfied for $Hr_i<1$.
  }
\label{fig:island2plot2}
\end{figure}

 As $c$ increases in $t$,  $(c G^{(3)}H)/(6\pi)$ gets close to ${\cal O}(1)$ such that $S_{\rm mat}(\Sigma) \gtrsim S_{\rm dS}$ is satisfied and  the information paradox arises.
 Then the region  $ Ht_i \lesssim 1$ begins to be allowed by condition 1.
 Moreover, the only negative term $-(1-H^2r_i^2)[(3\pi)/(cG^{(3)}H)]$ in LHS of the first inequality in \eqref{eq:T2Cond2} is no longer enhanced so $\partial_U S_{\rm mat}(I)\geq 0$ is easier to be satisfied.
 That is, when $(3\pi)/(cG^{(3)}H)$ is suppressed by the exponentially large $c$, the inequality reads
 \dis{Hr_i \gtrsim -\tanh[Ht_i]-\frac{1}{cH}\frac{dc}{dt_i}-\frac{3 dc/dt_i}{c^2H}S_{\rm mat}(I),}
which is trivially satisfied as RHS is negative.
 At the same time,  $\partial_V S_{\rm mat}(I)\leq 0$ in condition 2 becomes
 \dis{Hr_i \gtrsim \tanh[Ht_i]+\frac{1}{cH}\frac{dc}{dt_i}+\frac{3 dc/dt_i}{c^2H}S_{\rm mat}(I),\label{eq:Hrcond2}}
 which prefers the region $Ht_i < 1$.
 If $dc/dt_i$ were negligibly small, since the region $Ht_i\lesssim  1$ is now allowed by condition 1, we can find the region satisfying condition 1 and two inequalities in condition 2 simultaneously.
 Moreover, two inequalities in condition 3 are equivalent and trivially satisfied even if $(c G^{(3)}H)/(6\pi) \gtrsim 1$, so the island seems to be allowed  as shown in the left panel of Fig. \ref{fig:island2plot2}.
 However, we need to note here that $t_i$ for nonzero $dc/dt_i$ is interpreted as a time interval taken by the central charge to have the value  $c(t_i)$ through the time evolution.
 So the statement that $c$ is so large to satisfy $S_{\rm mat}(\Sigma)\simeq S_{\rm dS}$ at $H t_i\simeq 0$ means that $c$ is given by this large value from beginning, contradict to our setup.
 As can be inferred from \eqref{eq:timescale}, when $dc/dt_i$ is  close to zero (either $\gamma$ or $\epsilon_H$ is close to zero), $t_i$ becomes infinitely large, which excludes $Ht_i\simeq 0$.

 Moreover, when the increase of $(c G^{(3)}H)/(6\pi)$ is driven by the sizeable  positive $dc/dt$,  \eqref{eq:Hrcond2} is  more difficult to be satisfied.
 In this case, the inequality $\partial_V S_{\rm mat}(I)\leq 0$ is satisfied around  $Ht_i\simeq 0$ and $Hr_i\simeq 1$, in which $S_{\rm mat}(I)$ has a  small value.  
 Even if $t_i$ is negative, for $H|t_i|\gg 1$, $\tanh(Ht_i) \simeq 1$ but $S_{\rm mat}(I)$ is enhanced linear in $H|t_i|$ so $Ht_i\simeq 0$ is preferred by $\partial_V S_{\rm mat}(I)\leq 0$ as well.
 However, the region $Ht_i\simeq 0$ and $Hr_i\simeq 1$ is excluded by $\partial_U S_{\rm gen}(G)\leq 0$ in condition 3, as can be found in the first inequality in \eqref{eq:T2Cond3}.
 That is,  whereas $[(3dc/dt_i)/(c^2H)]S_{\rm mat}(I)$ is positive, other terms in LHS are suppressed for $Hr_i \simeq 1$ so LHS becomes positive, violating the inequality.
 This shows that   type 2 island is difficult to exist in the presence of the information paradox as well.
 Our discussion is summarized in the right panel of Fig. \ref{fig:island2plot2}.
 
 We also note that region $G$ is nothing more than $(\Sigma \cup I)^c$ thus one finds
 \dis{S_{\rm gen}(\Sigma \cup I)=\frac{\pi r}{G^{(3)}}+2 \frac{\pi r_i}{G^{(3)}}+c+\frac{c}{6}\log\Big[\frac{2}{\epsilon_{\rm UV}^2H^2}\big(1-\sqrt{1-H^2r_i^2}\big)\Big].}
 Since it depends on $t_i$ through $c$ satisfying $dc/dt_i >0$, $dS_{\rm gen}(\Sigma \cup I)/dt_i >0$ thus $S_{\rm gen}(\Sigma \cup I)$ is not extremized in $t_i$ direction. 
 Moreover, from
 \dis{\frac{S_{\rm gen}(\Sigma \cup I)}{dr_i}=\frac{2\pi}{G^{(3)}}+\frac{c}{6}\frac{H^2 r_i}{\sqrt{1-H^2 r_i^2}(1-\sqrt{1-H^2 r_i^2})} >0,}
 $S_{\rm gen}(\Sigma \cup I)$ is not extremized in $r_i$ direction.
 This shows that the island does not exist even if $c$ is constant in $t_i$.

\vskip 1.0cm

 From analysis of two types of island so far, we find that when the positive increasing rate of the central charge $c$  in time is the main reason to the increase of the radiation entropy in time, the information paradox arises but at the same time, $\partial_V S_{\rm gen}(I)\leq 0$ in condition 2 and  $\partial_U S_{\rm gen}(G)\leq 0$ in condition 3 become more restrictive. 
 We also check that the islands we considered cannot extremize $S_{\rm gen}(\Sigma \cup I)$ in a direct way.
 Even if we may find the parameter region allowing the island, we expect that this is a result of fine-tuning.


\section{Conclusion}

 In this article, we investigate when quasi-dS space  realizing cosmological inflation can have the information paradox and whether the island can exist without deforming the background far away from dS space,  emphasizing the connection between the thermal equilibrium and the dS isometries.
 For this purpose, we restrict the region collecting the radiation to the  static patch.
 Moreover, the radiation entropy is written in terms of the static coordinates such that it is independent of the static time coordinate in the perfect dS background, reflecting the equilibrium between   the radiation and the background.

 Our analysis shows that the information paradox in static patch can arise when the time evolution   of the radiation entropy $S_{\rm mat}(\Sigma)$   is dominated by the increase of the central charge.
 Since the central charge can be interpreted as the number of degrees of freedom, this is a reminiscent of the entropy argument supporting the dS swampland conjecture which states that the rapid descent of UV degrees of freedom violates the covariant entropy bound.
 If the island can exist within the  (quasi-)dS background, the radiation entropy  would not increase any longer, so the entropy argument does not work.
However,  the dominance of $dS_{\rm mat}(\Sigma)/dt$ by the increase of the central charge in time, nothing more than  the origin of the information paradox, also  prevents $S_{\rm gen}(I)$ and $S_{\rm gen}(G)$ from decreasing in one of the lightlike future direction.
 Then the island is difficult to exist in the quasi-dS background.
 Therefore, we can conclude that  the information paradox is not resolved within the (quasi-)dS space unless the parameters are fine-tuned, supporting the instability of  dS space as claimed by the dS swampland conjecture : for the central dogma to hold, spacetime needs to be strongly deformed from dS space.

As a possible way of   strong deformation of the background, we can consider the backreaction of  information collected by an observer, which breaks the dS isometries.
If the huge amount of collected information is concentrated on  the small region, black hole may be formed.
Then as considered in literatures discussing the island in dS space \cite{Chen:2020tes, Hartman:2020khs, Balasubramanian:2020xqf, Kames-King:2021etp, Shaghoulian:2021cef, Teresi:2021qff}, the island is allowed in the region inside the black hole horizon.
On the other hand, since  the modulus  responsible for the positive vacuum energy has a slow-roll  potential, spacetime eventually becomes Minkowski space, in which the observer does not find the information paradox.
We note here that if $\epsilon_H$ is very tiny (less than $H^2 G$ in four-dimensional case), quantum fluctuation  can prevent the modulus from slow-roll  so some region of spacetime may keep inflating, realizing the eternal inflation \cite{Steinhardt:1982, Vilenkin:1983xq, Linde:1986fc, Linde:1986fd, Goncharov:1987ir} (For a review, see, e.g., \cite{Guth:2007ng}).
Intriguingly, the condition allowing the eternal inflation is equivalent to the condition that  black hole is formed from the quantum fluctuation \cite{Seo:2021bpb} (See also \cite{Cohen:1998zx, Banks:2019arz}).
This is not strange since for black hole to be formed in dS space, the quantum fluctuation must be large enough  to produce the large mass concentration overwhelming the accelerating expansion of spacetime.
While both the black hole formation  and the slow roll with a sizeable $\epsilon_H$ (but still much smaller than $1$) can be ways to resolve the information paradox through the deformation of the background far away from dS space, these two have  different nature since in four-dimensional spacetime, $\epsilon_H$ is smaller than  $H^2G$ for the former and larger than $H^2G$ for the latter, respectively.
That is, the  unstable quasi-dS space can evolve into two different phases having different thermodynamic properties, depending on the size of the slow-roll parameter.


%

%


\appendix

\section{Appendix A : Modification of the action for time dependent $H$}
\label{app:modS}

Here we find the two-dimensional action for a $\tau$ dependent $H$ from the dimensional reduction. 
Since $H$ is no longer a constant in this case, $\Phi$ in  \eqref{eq:curvatures} is replaced by $\Phi/H(\tau)$ as the three-dimensional metric \eqref{eq:metric3to2} depends on $\Phi$ through the combination $\Phi/(2\pi H)$.
If we restrict our attention to the case of $g_{i\tau}=0=g^{i\tau}$  we obtain
\dis{R^{(3)}&=R^{(2)}-\frac{2H}{\Phi}\square^{(2)}\frac{\Phi}{H}
\\
&=R^{(2)}-\frac{2}{\Phi}\square^{(2)}\Phi-\frac{2H}{\Phi}g^{\tau\tau}\frac{d}{d\tau}(\epsilon_H \Phi)-\epsilon_H\frac{2H}{\Phi}\frac{1}{\sqrt{-g^{(2)}}}\partial_\tau(\sqrt{-g^{(2)}}g^{\tau\tau}\Phi),}
where $\epsilon_H=-H^{-2}dH/d\tau =d H^{-1}/d\tau$ as usually defined.
Combining this with $\sqrt{-g^{(3)}}=\sqrt{-g^{(2)}}[\Phi/(2\pi H)]$ and integrating over $\varphi$, we obtain the dimensional reduction of the Einstein-Hilbert term,
\dis{16\pi G^{(3)} S_{\rm EH}=\int d^2x \sqrt{-g^{(2)}}  \Big(\frac{\Phi}{H}R^{(2)}-\frac{2}{H}\square^{(2)}\Phi-2g^{\tau\tau}\frac{d}{d\tau}(\epsilon_H \Phi)-\frac{2\epsilon_H}{\sqrt{-g^{(2)}}}\partial_\tau(\sqrt{-g^{(2)}}g^{\tau\tau}\Phi )\Big).}
The first term is what we can find in \eqref{eq:JTaction}. 
 The second term and the last term can be rewritten as 
\dis{-\frac{2}{H}\sqrt{-g^{(2)}}\square^{(2)}\Phi=2\epsilon_H \sqrt{-g^{(2)}}g^{\tau\tau}\frac{d\Phi}{d\tau} -\partial_i\Big(\frac{2}{H}\sqrt{-g^{(2)}}g^{ij}\partial_j \Phi\Big), }
and
\dis{-2\epsilon_H \partial_\tau(\sqrt{-g^{(2)}}g^{\tau\tau}\Phi )= 2\frac{d\epsilon_H}{d\tau} \sqrt{-g^{(2)}} g^{\tau\tau}\Phi -\partial_\tau(2\epsilon_H\sqrt{-g^{(2)}} g^{\tau\tau}\Phi),}
respectively.
 Ignoring the surface term, the addition of these two terms gives
\dis{2\sqrt{-g^{(2)}}g^{\tau\tau}\frac{d}{d\tau}(\epsilon_H \Phi),}
which exactly cancels the third term in $S_{\rm EH}$.
Therefore, the dimensional reduction of the $S_{\rm EH}$ is simply given by
\dis{S_{\rm EH}=\int d^2x \sqrt{-g^{(2)}}\frac{\Phi}{16\pi G^{(3)} H} R^{(2)}.}
Since $H$ now varies with respect to $\tau$, the Newton's constant $G^{(3)}H$ is no longer a constant, even though it does not vary much over the time scale we consider.
Indeed, since we are interested in ${\cal O}(\epsilon_H^{1/2})$ corrections, the change of the Newton's constant of ${\cal O}(\epsilon_H)$ does not affect our analysis.
Meanwhile, in the flat coordinates, the dilaton can be approximated as $\Phi \simeq 2\pi H\rho e^{H\tau}$ if $\epsilon_H \ll 1$.
Then it is suggestive to introduce some fiducial constant $H_0$ such that we take a constant $G^{(3)}H_0$ to be a Newton's constant $G$ in which case the combination $\Phi/(G^{(3)}H)$ becomes $(H_0/H)\Phi/G$.
Then $(H_0/H)\Phi$ with $\Phi=2\pi H \rho a(\tau)$ depends on $\tau$ only through the scale factor $a(\tau)$, which was $e^{H\tau}$ for a constant $H$.
Since the cosmological constant term is not affected by the $\tau$ dependence of $H$, the bulk action is the same form as that given by \eqref{eq:JTaction}, which is  just a replacement of a constant $H$ by a $\tau$ dependent   $H(\tau)$.
Since the hypersurface orthogonal to $\tau$ direction has a rotation as an isometry, the FRW metric with the scale factor $a(\tau)$ satisfying  $H(\tau)=a^{-1}da/d\tau$ is expected to solve the equations of motion.

\renewcommand{\theequation}{\Alph{section}.\arabic{equation}}



\begin{thebibliography}{99}

\bibitem{Hawking:1976ra}
S.~W.~Hawking,
Phys. Rev. D \textbf{14} (1976), 2460-2473

\bibitem{Penington:2019npb}
G.~Penington,
JHEP \textbf{09} (2020), 002
[arXiv:1905.08255 [hep-th]].

\bibitem{Almheiri:2019psf}
A.~Almheiri, N.~Engelhardt, D.~Marolf and H.~Maxfield, 
JHEP \textbf{12} (2019), 063 
[arXiv:1905.08762 [hep-th]].


\bibitem{Almheiri:2019hni}
A.~Almheiri, R.~Mahajan, J.~Maldacena and Y.~Zhao,
JHEP \textbf{03} (2020), 149
[arXiv:1908.10996 [hep-th]].

\bibitem{Penington:2019kki}
G.~Penington, S.~H.~Shenker, D.~Stanford and Z.~Yang,
[arXiv:1911.11977 [hep-th]].

\bibitem{Almheiri:2019qdq}
A.~Almheiri, T.~Hartman, J.~Maldacena, E.~Shaghoulian and A.~Tajdini,
JHEP \textbf{05} (2020), 013
[arXiv:1911.12333 [hep-th]].

\bibitem{Almheiri:2020cfm}
A.~Almheiri, T.~Hartman, J.~Maldacena, E.~Shaghoulian and A.~Tajdini,
Rev. Mod. Phys. \textbf{93} (2021) no.3, 035002
[arXiv:2006.06872 [hep-th]].

\bibitem{Raju:2020smc}
S.~Raju,
Phys. Rept. \textbf{943} (2022), 1-80
[arXiv:2012.05770 [hep-th]].

\bibitem{Page:1993wv}
D.~N.~Page,
Phys. Rev. Lett. \textbf{71} (1993), 3743-3746
[arXiv:hep-th/9306083 [hep-th]].

\bibitem{Page:2013dx}
D.~N.~Page,
JCAP \textbf{09} (2013), 028
[arXiv:1301.4995 [hep-th]].

\bibitem{Chen:2020tes}
Y.~Chen, V.~Gorbenko and J.~Maldacena,
JHEP \textbf{02} (2021), 009
[arXiv:2007.16091 [hep-th]].

\bibitem{Hartman:2020khs}
T.~Hartman, Y.~Jiang and E.~Shaghoulian,
JHEP \textbf{11} (2020), 111
[arXiv:2008.01022 [hep-th]].

\bibitem{Balasubramanian:2020xqf}
V.~Balasubramanian, A.~Kar and T.~Ugajin,
JHEP \textbf{02} (2021), 072
[arXiv:2008.05275 [hep-th]].

\bibitem{Sybesma:2020fxg}
W.~Sybesma,
Class. Quant. Grav. \textbf{38} (2021) no.14, 145012
[arXiv:2008.07994 [hep-th]].

\bibitem{Geng:2021wcq}
H.~Geng, Y.~Nomura and H.~Y.~Sun,
Phys. Rev. D \textbf{103} (2021) no.12, 126004
[arXiv:2103.07477 [hep-th]].

\bibitem{Aalsma:2021bit}
L.~Aalsma and W.~Sybesma,
JHEP \textbf{05} (2021), 291
[arXiv:2104.00006 [hep-th]].

\bibitem{Kames-King:2021etp}
J.~Kames-King, E.~Verheijden and E.~Verlinde,
[arXiv:2108.09318 [hep-th]].

\bibitem{Shaghoulian:2021cef}
E.~Shaghoulian,
JHEP \textbf{01} (2022), 132
[arXiv:2110.13210 [hep-th]].


\bibitem{Teresi:2021qff}
D.~Teresi,
[arXiv:2112.03922 [hep-th]].

\bibitem{Bousso:2022gth}
R.~Bousso and E.~Wildenhain,
Phys. Rev. D \textbf{105} (2022) no.8, 086012
[arXiv:2202.05278 [hep-th]].



\bibitem{Espindola:2022fqb}
R.~Esp\'\i{}ndola, B.~Najian and D.~Nikolakopoulou,
[arXiv:2203.04433 [hep-th]].

\bibitem{Gibbons:1977mu}
G.~W.~Gibbons and S.~W.~Hawking,
Phys. Rev. D \textbf{15} (1977), 2738-2751

\bibitem{Jacobson:2003wv}
T.~Jacobson and R.~Parentani,
Found. Phys. \textbf{33} (2003), 323-348
[arXiv:gr-qc/0302099 [gr-qc]].

\bibitem{Unruh:1976db}
W.~G.~Unruh,
Phys. Rev. D \textbf{14} (1976), 870

\bibitem{Hartle:1976tp}
J.~B.~Hartle and S.~W.~Hawking,
Phys. Rev. D \textbf{13} (1976), 2188-2203

\bibitem{Chernikov:1968zm}
N.~A.~Chernikov and E.~A.~Tagirov,
Ann. Inst. H. Poincare Phys. Theor. A \textbf{9} (1968), 109

\bibitem{Bunch:1978yq}
T.~S.~Bunch and P.~C.~W.~Davies,
Proc. Roy. Soc. Lond. A \textbf{360} (1978), 117-134



\bibitem{Aalsma:2019rpt}
L.~Aalsma, M.~Parikh and J.~P.~Van Der Schaar,
JHEP \textbf{11} (2019), 136
[arXiv:1905.02714 [hep-th]].

\bibitem{Gong:2020mbn}
J.~O.~Gong and M.~S.~Seo,
JCAP \textbf{10} (2021), 042
[arXiv:2011.01794 [hep-th]].

\bibitem{Teitelboim:1983ux}
C.~Teitelboim,
Phys. Lett. B \textbf{126} (1983), 41-45

\bibitem{Jackiw:1984je}
R.~Jackiw,
Nucl. Phys. B \textbf{252} (1985), 343-356






\bibitem{Obied:2018sgi}
G.~Obied, H.~Ooguri, L.~Spodyneiko and C.~Vafa,
[arXiv:1806.08362 [hep-th]].

\bibitem{Andriot:2018wzk}
D.~Andriot,
Phys. Lett. B \textbf{785} (2018), 570-573
[arXiv:1806.10999 [hep-th]].

\bibitem{Garg:2018reu}
S.~K.~Garg and C.~Krishnan,
JHEP \textbf{11} (2019), 075
[arXiv:1807.05193 [hep-th]].

\bibitem{Ooguri:2018wrx}
H.~Ooguri, E.~Palti, G.~Shiu and C.~Vafa,
Phys. Lett. B \textbf{788} (2019), 180-184
[arXiv:1810.05506 [hep-th]].

\bibitem{Ooguri:2006in}
H.~Ooguri and C.~Vafa,
Nucl. Phys. B \textbf{766} (2007), 21-33
[arXiv:hep-th/0605264 [hep-th]].

\bibitem{Etheredge:2022opl}
M.~Etheredge, B.~Heidenreich, S.~Kaya, Y.~Qiu and T.~Rudelius,
[arXiv:2206.04063 [hep-th]].

\bibitem{Zamolodchikov:1986gt}
A.~B.~Zamolodchikov,
JETP Lett. \textbf{43} (1986), 730-732




\bibitem{Seo:2019mfk}
M.~S.~Seo,
Phys. Lett. B \textbf{797} (2019), 134904
[arXiv:1907.12142 [hep-th]].

\bibitem{Seo:2019wsh}
M.~S.~Seo,
Phys. Lett. B \textbf{807} (2020), 135580
[arXiv:1911.06441 [hep-th]].

\bibitem{Sun:2019obt}
S.~Sun and Y.~L.~Zhang,
Phys. Lett. B \textbf{816} (2021), 136245
[arXiv:1912.13509 [hep-th]].

\bibitem{Aguilar-Gutierrez:2021bns}
S.~E.~Aguilar-Gutierrez, A.~Chatwin-Davies, T.~Hertog, N.~Pinzani-Fokeeva and B.~Robinson,
JHEP \textbf{11} (2021), 212
[arXiv:2108.01278 [hep-th]].

\bibitem{Holzhey:1994we}
C.~Holzhey, F.~Larsen and F.~Wilczek,
Nucl. Phys. B \textbf{424} (1994), 443-467
[arXiv:hep-th/9403108 [hep-th]].

\bibitem{Calabrese:2004eu}
P.~Calabrese and J.~L.~Cardy,
J. Stat. Mech. \textbf{0406} (2004), P06002
[arXiv:hep-th/0405152 [hep-th]].

\bibitem{Calabrese:2009qy}
P.~Calabrese and J.~Cardy,
J. Phys. A \textbf{42} (2009), 504005
[arXiv:0905.4013 [cond-mat.stat-mech]].


\bibitem{Strominger:2001pn}
A.~Strominger,
JHEP \textbf{10} (2001), 034
[arXiv:hep-th/0106113 [hep-th]].

\bibitem{Hu:1986jd}
B.~L.~Hu and D.~Pavon,
Phys. Lett. B \textbf{180} (1986), 329-334

\bibitem{Brandenberger:1992sr}
R.~H.~Brandenberger, V.~F.~Mukhanov and T.~Prokopec,
Phys. Rev. Lett. \textbf{69} (1992), 3606-3609
[arXiv:astro-ph/9206005 [astro-ph]].

\bibitem{Brandenberger:1992jh}
R.~H.~Brandenberger, T.~Prokopec and V.~F.~Mukhanov,
Phys. Rev. D \textbf{48} (1993), 2443-2455
[arXiv:gr-qc/9208009 [gr-qc]].

\bibitem{Prokopec:1992ia}
T.~Prokopec,
Class. Quant. Grav. \textbf{10} (1993), 2295-2306

\bibitem{Gasperini:1992xv}
M.~Gasperini and M.~Giovannini,
Phys. Lett. B \textbf{301} (1993), 334-338
[arXiv:gr-qc/9301010 [gr-qc]].

\bibitem{Gasperini:1993mq}
M.~Gasperini and M.~Giovannini,
Class. Quant. Grav. \textbf{10} (1993), L133-L136
[arXiv:gr-qc/9307024 [gr-qc]].

\bibitem{Brahma:2020zpk}
S.~Brahma, O.~Alaryani and R.~Brandenberger,
Phys. Rev. D \textbf{102} (2020) no.4, 043529
[arXiv:2005.09688 [hep-th]].




\bibitem{Cai:2019dzj}
R.~G.~Cai and S.~J.~Wang,
Sci. China Phys. Mech. Astron. \textbf{64} (2021) no.1, 210011
[arXiv:1912.00607 [hep-th]].

\bibitem{Saha:2021ohr}
A.~Saha, S.~Gangopadhyay and J.~P.~Saha,
[arXiv:2109.02996 [hep-th]].

\bibitem{Steinhardt:1982}
P.~J.~Steinhardt, ``Natural inflation," in The Very Early
Universe, Proceedings of the Nuffield Workshop, Cambridge, 21 June - 9 July, 1982, eds: G.~W.~Gibbons, S.~W.~Hawking and S.~T.~C.~ Siklos (Cambridge University Press).



\bibitem{Vilenkin:1983xq}
A.~Vilenkin,
Phys. Rev. D \textbf{27} (1983), 2848

\bibitem{Linde:1986fc}
A.~D.~Linde,
Mod. Phys. Lett. A \textbf{1} (1986), 81

\bibitem{Linde:1986fd}
A.~D.~Linde,
Phys. Lett. B \textbf{175} (1986), 395-400

\bibitem{Goncharov:1987ir}
A.~S.~Goncharov, A.~D.~Linde and V.~F.~Mukhanov,
Int. J. Mod. Phys. A \textbf{2} (1987), 561-591

\bibitem{Guth:2007ng}
A.~H.~Guth,
J. Phys. A \textbf{40} (2007), 6811-6826
[arXiv:hep-th/0702178 [hep-th]].

\bibitem{Seo:2021bpb}
M.~S.~Seo,
[arXiv:2106.00138 [hep-th]].

\bibitem{Cohen:1998zx}
A.~G.~Cohen, D.~B.~Kaplan and A.~E.~Nelson,
Phys. Rev. Lett. \textbf{82} (1999), 4971-4974
[arXiv:hep-th/9803132 [hep-th]].

\bibitem{Banks:2019arz}
T.~Banks and P.~Draper,
Phys. Rev. D \textbf{101} (2020) no.12, 126010
[arXiv:1911.05778 [hep-th]].



\end{thebibliography}
\end{document}